\documentclass[12pt,draftclsnofoot,onecolumn,romanappendices]{IEEEtran}

\IEEEoverridecommandlockouts

\usepackage{algorithm,algorithmic,amsmath,amssymb,amsthm,array,bbm,bibentry,cite,color,comment}
\usepackage{enumerate,enumitem,eurosym,float,graphicx,lettrine,mathrsfs,microtype,multirow,nomencl}
\usepackage{pict2e,psfrag,ragged2e,setspace,stfloats,subfigure,supertabular,tabularx,url}
\usepackage[USenglish]{babel}

{		\newtheorem{theorem}{Theorem}}
{ 		}
{ 		\newtheorem{lemma}{Lemma}}
{ 		\newtheorem{proposition}{Proposition}}
{ 		\newtheorem{remark}{Remark}}
{		}
{ 		\newtheorem{corollary}{Corollary}}

\newcommand{\rmR}{\mathrm{R}}
\newcommand{\rmT}{\mathrm{T}}
\newcommand{\SINR}{\mathrm{SINR}}

\newcommand{\Psuc}{\mathrm{P}_{\mathrm{suc}}}
\newcommand{\Psuclb}{\underline{\mathrm{P}}_{\mathrm{suc}}}

\newcommand\narrowstyle{\SetTracking{encoding=*}{-1}\lsstyle}
\newcommand{\red}[1]{\textcolor{black}{#1}}

\newcommand{\h}{\mathbf{h}}

\newcommand{\n}{\mathbf{n}}

\renewcommand{\v}{\mathbf{v}}
\newcommand{\w}{\mathbf{w}}

\newcommand{\y}{\mathbf{y}}

\newcommand{\E}{\mathbf{E}}

\renewcommand{\H}{\mathbf{H}}
\newcommand{\I}{\mathbf{I}}

\newcommand{\Q}{\mathbf{Q}}

\newcommand{\Sigmab}{\mathbf{\Sigma}}

\newcommand{\setC}{\mathcal{C}}

\newcommand{\setL}{\mathcal{L}}

\newcommand{\setN}{\mathcal{N}}

\newcommand{\Real}{\mbox{$\mathbb{R}$}}
\newcommand{\Compl}{\mbox{$\mathbb{C}$}}

\newcommand{\diag}{\mathrm{diag}}

\newcommand{\diff}{\mathrm{d}}
\newcommand{\Exp}{\mathbb{E}}
\newcommand{\herm}{\mathrm{H}}

\renewcommand{\Pr}{\mathbb{P}}

\newcommand{\tr}{\mathrm{tr}}

\newcommand{\Var}{\mathrm{Var}}

\title{Full-Duplex MIMO Small-Cell Networks \\ with Interference Cancellation}

\author{Italo Atzeni, \textit{Member}, \textit{IEEE}, and Marios Kountouris, \textit{Senior Member}, \textit{IEEE} \thanks{Part of this work was presented at GLOBECOM 2015 in San Diego, CA, USA \cite{Atz15a} and at Asilomar 2016 in Pacific Grove, CA, USA \cite{Atz16b}.} \\ \thanks{The authors are with the Mathematical and Algorithmic Sciences Lab, Paris Research Center, Huawei Technologies Co.,~Ltd., France (email: italo.atzeni@huawei.com; marios.kountouris@huawei.com).}}

\makeindex

\begin{document}

\maketitle

\begin{abstract}
Full-duplex (FD) technology is envisaged as a key component for future mobile broadband networks due to its ability to boost the spectral efficiency. FD systems can transmit and receive simultaneously on the same frequency at the expense of residual self-interference \red{(SI)} and additional interference to the network compared with half-duplex (HD) transmission. This paper analyzes the performance of wireless networks with FD multi-antenna base stations (BSs) and HD user equipments (UEs) using stochastic geometry. Our analytical results quantify the success probability and the achievable spectral efficiency and indicate the amount of \red{SI} cancellation needed for beneficial FD operation. \red{The advantages of multi-antenna BSs/UEs are shown and the performance gains achieved by balancing desired signal power increase and interference cancellation are derived.} \red{The proposed framework aims at shedding light on the system-level gains of FD mode with respect to HD mode in terms of network throughput, and provides design guidelines for the practical implementation of FD technology in large small-cell networks.}
\end{abstract}


\section{Introduction} \label{sec:Intro}

Full-duplex (FD) communication is an emerging technology that has been recognized as one of the promising solutions to cope with the \red{growing} demand for high data rates. Indeed, allowing the network nodes to transmit and receive over the same time/frequency resources can potentially double spectral efficiency with respect to the half-duplex (HD) counterparts (i.e., time- and frequency-division duplex) \cite{Sab14}. However, there are three major technical challenges hindering the implementation of FD cellular networks. \red{First, the signal reception is affected by the \textit{self-interference} \red{(SI)}, i.e., the signal leakage resulting from the imperfect isolation between transmit and receive antennas \cite{Dua12}. Second, \textit{inter-node} interference arises due to the simultaneous uplink (UL)/downlink (DL) communications of nodes in the same cell \cite{Ale16}.} Third, the concurrent, aggressive utilization of both forward and reverse links doubles the interference between neighboring cells \cite{Goy15}.

Recently, there has also been an increasing interest in network densification as a means to fulfill the performance requirements of 5th generation (5G) wireless systems \cite{Bhu14}. In particular, ultra-dense networks (UDNs), i.e., the dense and massive deployment of small-cell base stations (BSs), is regarded as a key enabler for providing higher data rates and enhanced coverage by exploiting spatial reuse. Interestingly, small-cell BSs prove particularly suitable for the deployment of FD technology thanks to their reduced transmit power and the low mobility of their user equipments (UEs). \red{In this respect, the hybrid FD/HD network configuration, with small-cell BSs operating in FD mode and UL/DL nodes operating in HD mode, is appropriate for UDN scenarios since it exploits the throughput gains promised by FD at the BSs while reducing the overall interference of the system \cite{Goy15,Sha17}.}
This hybrid FD/HD architecture can be used either for serving an UL node and a DL node separately (with two independent data flows) or for relaying purposes to increase coverage between an UL node and a DL node (with the same data flow being received, amplified, and re-transmitted by the FD BS) \cite{Sab14}.

\subsection{Related Work} \label{sec:Intro_SOTA}

\begin{figure}[t!]
\centering
\includegraphics[scale=0.95]{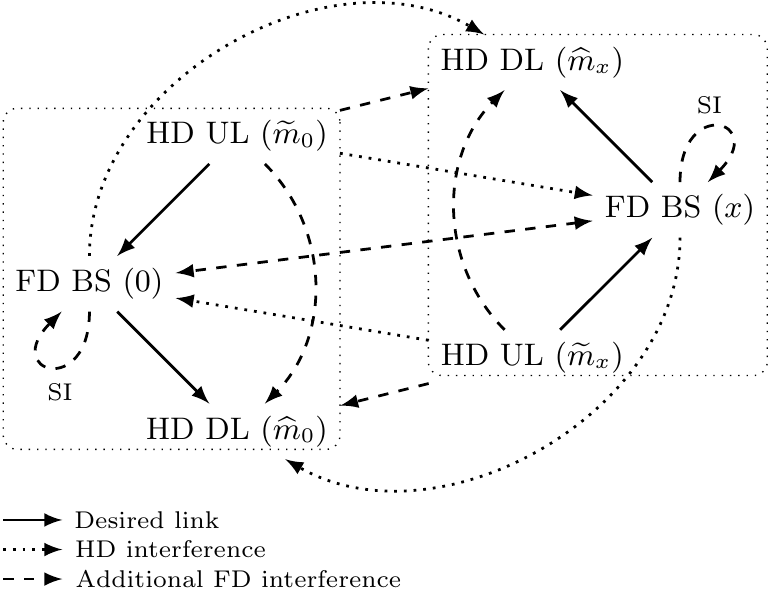} 
\caption{System model with FD BSs relaying between HD UL and DL nodes, with corresponding desired signals and interferences.} \label{fig:topology} \vspace{-3mm}
\end{figure}

Due to the extra interference terms introduced in FD mode (see Figure~\ref{fig:topology}), it is not clear how the network throughput and the aggregate interference will behave in dense multi-cell FD systems. Several recent works have examined the performance of large-scale FD networks using stochastic geometry, which is a powerful mathematical framework that provides models and tools for efficiently analyzing the performance of UDNs and heterogeneous cellular/ad hoc~networks~\cite{Hae12}.

References \cite{Ton15,Lee15} study the performance of bipolar networks and multi-tier heterogeneous networks, respectively, consisting of both HD and FD nodes, and quantify the impact of imperfect \red{SI} cancellation. Interestingly, the two papers reach the same conclusion that operating all nodes in either FD or HD mode maximizes the area spectral efficiency compared with a mixture of the two modes. On the other hand, \cite{Goy16} shows that raising the proportion of FD nodes increases the outage probability and thus highlights the inherent tradeoff between coverage and throughput. \red{Furthermore, \cite{Sha17} analyzes the DL performance of FD self-backhauling small cells in a two-tier heterogeneous network, showing that the rate could be close to double that of a conventional HD self-backhauling network at the expense of reduced coverage.} All these works assume single-antenna nodes, whereas multiple-input multiple-output (MIMO) nodes are considered by \cite{Moh15} in a single-cell setting with randomly located DL UEs. Furthermore, \cite{Pso16} analyzes the effect of directional antennas in reducing the overall interference in FD cellular networks and shows that directionality alone can passively suppress the \red{SI}. The asymptotic performance of massive MIMO-enabled backhaul nodes serving FD small-cell BSs is studied in \cite{Tab16}, where zero-forcing beamforming allows to mitigate the interference among multiple backhaul data streams. The scenario of \red{FD BSs} with massive antenna arrays is considered in \cite{Sho16}, which also assumes mobile UEs with FD capabilities.

\subsection{Motivation and Contributions} \label{sec:Intro_contr}

{\narrowstyle\red{The performance of FD technology with multiple antennas has not been studied in large-scale systems, if not asymptotically for massive MIMO.} Nevertheless, while massive antenna arrays are appropriate for macro-cell backhaul nodes, they are not suitable for small-cell BSs typical of UDN scenarios, which are usually equipped with a low-to-moderate number of transmit/receive antennas. Hence, it is meaningful to investigate the performance of dense FD small-cell networks with multi-antenna nodes, which is a promising and practically relevant solution for future mobile broadband networks. In particular, it is important to analyze the impact of array gain and interference cancellation techniques in mitigating the additional interference introduced by FD mode.}

\red{On the other hand, assuming that the \red{SI} channel is subject to Rayleigh fading or adopting a constant value to model its power gain have been common practices in the literature (with the exception of our previous work \cite{Atz15a} and the very recent paper \cite{Sho17})}. However, these are very coarse approximations: the former neglects the strong line-of-sight signal component between transmit and receive antennas \cite{Dua12}, whereas the latter is only meaningful when digital cancellation is applied \cite{Rii11}. As a matter of fact, the residual \red{SI} channel is known to be subject to Rician fading\footnote{Before applying active cancellation, the magnitude of the \red{SI} channel is modeled using a Rician distribution with large $K$-factor due to the strong line-of-sight component; after applying active cancellation, the line-of-sight component is reduced, resulting in smaller $K$-factor \cite{Dua12}.} and, therefore, its modeling in a MIMO context represents a challenging problem when receive combining and transmit beamforming techniques are employed.

\red{Prior work fails to unveil the real potential of MIMO techniques in large-scale FD systems, for which a precise characterization of the \red{SI} power is crucial. In this paper, we fill these gaps by providing the following contributions:}
\begin{itemize}
\item[$\bullet$] Using tools from stochastic geometry, we study the performance of wireless networks with randomly distributed FD MIMO relays and derive tight bounds for the probability of successful transmission. \red{The proposed framework can be used to shed light on the system-level gains of FD mode with respect to HD mode in terms of network throughput.}
\item[$\bullet$] We appropriately model the \red{SI} channel using Rician fading and we derive the distribution of the \red{SI} power for arbitrary receive combining and transmit beamforming strategies. The resulting expression approximately follows a gamma distribution and can be readily incorporated into existing frameworks for the performance analysis of UDNs.
\item[$\bullet$] We consider interference cancellation techniques at the receive side of both the \red{FD BSs} and the (multi-antenna) DL nodes. \red{In particular, we analyze different receive configurations based on partial zero forcing (PZF), which allows to identify which interference terms are most critical for the deployment of FD technology and their interplay.}
\end{itemize}

The rest of the paper is structured as follows. The system model is introduced in Section~\ref{sec:SM}. Section~\ref{sec:SP} presents our main results on the performance analysis of FD MIMO small-cell networks. The proposed analytical framework is extended in Section~\ref{sec:IC} to accommodate interference cancellation techniques at both the FD BSs and the HD DL nodes. In Section~\ref{sec:NUM}, numerical results are reported to corroborate our theoretical findings and to establish under which conditions FD mode outperforms HD mode. Finally, Section~\ref{sec:END} summarizes our contributions and draws some concluding remarks.

\section{System Model} \label{sec:SM}
\subsection{Network Model} \label{sec:SM_NS}

We consider a set of small-cell BSs operating in FD mode. Each FD BS acts as a relay between an UL node and a DL node,\footnote{We assume that the FD BSs adopt a perfect decode-and-forward relaying scheme; the study of imperfect schemes goes beyond the scope of this paper.} both operating in HD mode, during a given time slot; all communications occur in the same frequency band. This general scenario is depicted in Figure~\ref{fig:topology} and can be used to model, for instance, the two-hop communication between a backhaul node and a mobile UEs, or self-backhauled small cells. In our setting, the FD BSs are equipped with multiple receive/transmit antennas, \red{whereas the HD UL/DL nodes are assumed to have a single receive/transmit antenna for simplicity (as, e.g., in \cite{Sha17})}. Alternatively, our model can be seen as an instance of multi-antenna UL/DL where nodes perform space division multiple access (SDMA) and send/receive one stream to/from the FD BS, hence being equivalently seen as single-antenna nodes by each FD BS \red{\cite{Dhi13}}. In short, our model consists of a single-input multiple-output \red{(SIMO)} transmission followed by a multiple-input single-output \red{(MISO)} transmission.\footnote{\red{Observe that such network model where each FD BS is associated with a single-antenna UL node can be regarded as a tractable approximation of the more realistic case where several FD BSs are served by the same multi-antenna UL node, since the number of UL interfering transmissions seen by any FD BS or DL node is the same for both scenarios.}}

Let us thus introduce the stationary, independently marked Poisson point process (PPP) $\Phi_{\mathrm{m}} \triangleq \big\{ (x_{i}, \widetilde{m}(x_{i}), \widehat{m}(x_{i})) \big\}$ on $\Real^{2} \times \Real^{2} \times \Real^{2}$. \red{The ground process $\Phi \triangleq \{ x_{i} \}$, which includes the locations of the FD BSs, is a PPP with spatial density $\lambda$, measured in [BSs/m$^{2}$]. Likewise, $\widetilde{\Phi} \triangleq \widetilde{m}(\Phi) = \{ \widetilde{m}(x) \}_{x \in \Phi}$ and $\widehat{\Phi} \triangleq \widehat{m}(\Phi) = \{ \widehat{m}(x) \}_{x \in \Phi}$ are the isotropic marks of $\Phi$, which include the locations of the HD UL and DL nodes, respectively,} with fixed distances of the desired links given by $\widetilde{R} \triangleq \| x - \widetilde{m}(x) \|$ and $\widehat{R} \triangleq \| x - \widehat{m}(x) \|$, $\forall x \in \Phi$. Therefore, we have $\widetilde{m}(x) = x + \widetilde{R} (\cos \widetilde{\varphi}_{x}, \sin \widetilde{\varphi}_{x})$ and $\widehat{m}(x) = x + \widehat{R} (\cos \widehat{\varphi}_{x}, \sin \widehat{\varphi}_{x})$, with $\{ \widetilde{\varphi}_{x}, \widehat{\varphi}_{x} \}_{x \in \Phi}$ independent and uniformly distributed in $[0,2 \pi]$. \red{Evidently, $\widetilde{\Phi}$ and $\widehat{\Phi}$ are also PPPs with density $\lambda$ and are dependent on $\Phi$.} For convenience, in the rest of the paper we use the notation $\widetilde{m}_{x} \triangleq \widetilde{m}(x)$ and $\widehat{m}_{x} \triangleq \widehat{m}(x)$. \red{Observe that one can consider random distances of the links by first conditioning on $\widetilde{R}$ and $\widehat{R}$ and then averaging over $\widetilde{R}$ and $\widehat{R}$, without affecting the main conclusions of this paper; a similar network models with fixed distances between transmitters and receivers have been adopted, among others, in \cite{Ton15,Vaz11,Jin08}.}

\subsection{Channel Model} \label{sec:SM_CM}

We assume that the FD BSs and the HD UL nodes transmit with constant powers $\widehat{\rho}$ and $\widetilde{\rho}$, respectively. Furthermore, the FD BSs are equipped with $N_{\rmR}$ receive antennas and $N_{\rmT}$ transmit antennas.

The propagation through the wireless channel is characterized as the combination of a pathloss attenuation and a small-scale fading. Given transmitting node $x$ and receiving node $z$, we use the following notation. The pathloss between nodes $x$ and $z$ is given by the function $\ell (x,z) \triangleq \|x - z\|^{-\alpha}$, with pathloss exponent $\alpha > 2$. The channels are denoted by $\H_{x z} \in \Compl^{N_{\rmR} \times N_{\rmT}}$ if $x, z \in \Phi$, as $\h_{x z} \in \Compl^{N_{\rmR} \times 1}$ if $x \in \widetilde{\Phi}$ and $z \in \Phi$, as $\h_{x z} \in \Compl^{N_{\rmT} \times 1}$ if $x \in \Phi$ and $z \in \widehat{\Phi}$, and as $h_{x z} \in \Compl$ if $x \in \widetilde{\Phi}$ and $z \in \widehat{\Phi}$; in particular, $\H_{x x}$ models the \red{SI} at $x \in \Phi$ resulting from its own transmission. We assume that all the channels, except the \red{SI} channel, are subject to Rayleigh fading with elements distributed independently as $\setC \setN (0, 1)$. On the other hand, the \red{SI} channel is subject to Rician fading \cite{Dua12} and, therefore, \red{the elements of $\H_{x x}$ are distributed independently as $\setC \setN (\mu_{i j}, \nu^{2})$, where $\mu_{i j} \in \Compl$ is the mean of the $(i,j)$-th element (independent across elements), with the same absolute mean $\mu \triangleq | \mu_{i j} |$, $\forall i=1, \ldots, N_{\rmR}$, $\forall j=1, \ldots, N_{\rmT}$. In this regard, one can measure the Rician $K$-factor and the \red{SI} attenuation $\Omega$ between transmit and receive antennas and determine the absolute mean} and standard deviation of $\H_{x x}$ as (cf. \cite{Tep03})
\begin{align} \label{eq:mu_nu}
\mu \triangleq \sqrt{\frac{K \Omega}{K+1}}, \qquad \nu \triangleq \sqrt{\frac{\Omega}{K+1}}.
\end{align}

In addition, let $s_{x}$ represent the data symbol transmitted by $x$ with $\Exp [ |s_{x}|^2 ] = 1$, whereas the additive noise at $x$ is denoted by $\n_{x} \in \Compl^{N_{\rmR}}$ if $x \in \Phi$ and by $n_{x} \in \Compl$ if $x \in \widehat{\Phi}$, with elements distributed independently as $\setC \setN (0, \sigma^{2})$. Lastly, $\v_{x} \in \Compl^{N_{\rmR}}$ and $\w_{x} \in \Compl^{N_{\rmT}}$ denote the receive combining and the transmit beamforming vectors applied by $x \in \Phi$, respectively, with $\| \v_{x} \|^{2} = \| \w_{x} \|^{2} = 1$.

\subsection{SINR Characterization} \label{sec:SM_SINR}

In this section, we characterize the signal-to-interference-plus-noise ratio (SINR) at the FD BSs and at the HD DL nodes, which is then used to analyze the probability of successful transmission, also termed as \textit{success probability}, in the next section. \red{Here, we focus on the first hop (i.e., the SINR at the FD BSs) and the second hop (i.e., the SINR at the HD DL nodes) separately.} Our analysis focuses on a \textit{typical FD BS}, indexed by $0$, and on its corresponding HD UL/DL nodes, referred to as \textit{typical HD UL/DL nodes} and indexed by $\widetilde{m}_{0}$ and $\widehat{m}_{0}$, respectively. The two-hop link between these nodes is representative of the whole network, as detailed next.

\smallskip

\noindent \textbf{First Hop:} Consider the typical FD BS located at the origin of the Euclidean plane and indexed by $0$. Due to Slivnyak's theorem \cite[Ch.~8.5]{Hae12} and to the stationarity of $\Phi$, the statistics of the typical BS's signal reception are representative of the statistics seen by any FD BS: we can thus write $\ell (x,0) = r_{x}^{-\alpha}$, with $r_{x} \triangleq \| x \|$ being the distance of $x$ from the typical FD BS. Hence, the received signal at the typical FD BS is given by
\begin{align}
\label{eq:y_1} \y_{0} \triangleq & \ \underbrace{\sqrt{\widetilde{\rho}} \widetilde{R}^{-\frac{\alpha}{2}} \h_{\widetilde{m}_{0} 0} s_{\widetilde{m}_{0}}}_{\mathrm{(a)}} + \sum_{x \in \Phi} \underbrace{\sqrt{\widehat{\rho}} r_{x}^{-\frac{\alpha}{2}} \H_{x 0} \w_{x} s_{x}}_{\mathrm{(b)}} + \sum_{x \in \Phi} \underbrace{\sqrt{\widetilde{\rho}} r_{\widetilde{m}_{x}}^{-\frac{\alpha}{2}} \h_{\widetilde{m}_{x} 0} s_{\widetilde{m}_{x}}}_{\mathrm{(c)}} + \underbrace{\sqrt{\widehat{\rho}} \H_{0 0} \w_{0} s_{0}}_{\mathrm{(d)}} + \n_{0}
\end{align}
where (a) represents the desired signal, (b) and (c) indicate the interference coming from FD BS $x$ and its associated HD UL node $\widetilde{m}_{x}$, respectively, and (d) represents the \red{SI}. Given the receive combining vector $\v_{0}$, the resulting SINR reads as
\begin{align} \label{eq:SINR_1}
\SINR_{0} & \triangleq \frac{\widetilde{\rho} \widetilde{R}^{-\alpha} S_{\widetilde{m}_{0} 0}}{I_{0} + \sigma^{2}}
\end{align}
where we have defined
\begin{equation}
S_{x 0} \triangleq \left\{
\begin{array}{ll}
|\v_{0}^{\herm} \H_{x 0} \w_{x}|^{2}, & x \in \Phi \\
|\v_{0}^{\herm} \h_{x 0}|^{2}, & x \in (\widetilde{\Phi} \cup \widetilde{m}_{0})
\end{array} \right.
\end{equation}
and where $I_{0}$ is the overall interference power at the typical FD BS, i.e.,
\begin{align} \label{eq:I_1}
I_{0} & \triangleq \sum_{x \in \Phi} \big( \widehat{\rho} r_{x}^{-\alpha} S_{x 0} + \widetilde{\rho} r_{\widetilde{m}_{x}}^{-\alpha} S_{\widetilde{m}_{x} 0} \big) + \widehat{\rho} S_{0 0}.
\end{align}
The success probability of the first hop is derived in Section~\ref{sec:SP_1}.

\smallskip

\noindent \textbf{Second Hop:} \red{Consider the typical HD DL node located at distance $\widehat{R}$ from the origin of the Euclidean plane and indexed by $\widehat{m}_{0}$.} Again, following Slivnyak's theorem and due to the stationarity of $\widehat{\Phi}$, the statistics of the typical HD DL node's signal reception are representative of the statistics seen by any HD DL node: we can thus write $\ell (x,\widehat{m}_{0}) = r_{x}^{-\alpha}$, with $r_{x} \triangleq \| x - \widehat{m}_{0} \|$ being the distance of $x$ from the typical HD DL node. Hence, the received signal at the typical HD DL node is given by \vspace{-1mm}
\begin{align}
\nonumber & \hspace{-15mm} y_{\widehat{m}_{0}} \triangleq \ \underbrace{\sqrt{\widehat{\rho}} \widehat{R}^{-\frac{\alpha}{2}} \h_{0 \widehat{m}_{0}}^{\herm} \w_{0} s_{0}}_{\mathrm{(a)}} + \sum_{x \in \Phi} \underbrace{\sqrt{\widehat{\rho}} r_{x}^{-\frac{\alpha}{2}} \h_{x \widehat{m}_{0}}^{\herm} \w_{x} s_{x}}_{\mathrm{(b)}} \\
\label{eq:y_2} & \hspace{30mm} + \sum_{x \in \Phi} \underbrace{\sqrt{\widetilde{\rho}} r_{\widetilde{m}_{x}}^{-\frac{\alpha}{2}} h_{\widetilde{m}_{x} \widehat{m}_{0}} s_{\widetilde{m}_{x}}}_{\mathrm{(c)}} + \underbrace{\sqrt{\widetilde{\rho}} r_{\widetilde{m}_{0}}^{-\frac{\alpha}{2}} h_{\widetilde{m}_{0} \widehat{m}_{0}} s_{\widetilde{m}_{0}}}_{\mathrm{(d)}} + n_{\widehat{m}_{0}}
\end{align}
where (a) represents the desired signal, (b) and (c) indicate the interference coming from FD BS $x$ and its associated HD UL node $\widetilde{m}_{x}$, respectively, and (d) represents the inter-node interference \red{coming from the HD UL node $\widetilde{m}_{0}$ in the same cell}. The resulting SINR reads as \vspace{-2mm}
\begin{align} \label{eq:SINR_2}
\SINR_{\widehat{m}_{0}} & \triangleq \frac{\widehat{\rho} \widehat{R}^{-\alpha} S_{0 \widehat{m}_{0}}}{I_{\widehat{m}_{0}} + \sigma^{2}}
\end{align}
where we have defined
\begin{equation}
S_{x \widehat{m}_{0}} \triangleq \left\{
\begin{array}{ll}
|\h_{x \widehat{m}_{0}}^{\herm} \w_{x}|^{2}, & x \in (\Phi \cup 0) \\
|h_{x \widehat{m}_{0}}|^{2}, & x \in (\widetilde{\Phi} \cup \widetilde{m}_{0})
\end{array} \right.
\end{equation}
and where $I_{\widehat{m}_{0}}$ is the overall interference power at $\widehat{m}_{0}$, i.e., \vspace{-1mm}
\begin{align} \label{eq:I_2}
I_{\widehat{m}_{0}} \triangleq \sum_{x \in \Phi} \big( \widehat{\rho} r_{x}^{-\alpha} S_{x \widehat{m}_{0}} + \widetilde{\rho} r_{\widetilde{m}_{x}}^{-\alpha} S_{\widetilde{m}_{x} \widehat{m}_{0}} \big) + \widetilde{\rho} r_{\widetilde{m}_{0}}^{-\alpha} S_{\widetilde{m}_{0} \widehat{m}_{0}}.
\end{align}
The success probability of the second hop is derived in Section~\ref{sec:SP_2}.

For the sake of simplicity, we focus on the interference-limited case, where $I_{0} \gg \sigma^{2}$ and $I_{\widehat{m}_{0}} \gg \sigma^{2}$, and consider the signal-to-interference ratio (SIR). Our analysis can be extended with straightforward yet more involved calculations to the general case.

\section{Success Probability} \label{sec:SP}

\red{The successful transmission of a packet over the complete communication path, i.e., from the HD UL node to the HD DL node through the FD BS, is given by the joint complementary cumulative distribution function (CCDF) of $\SINR_{0}$ and $\SINR_{\widehat{m}_{0}}$, which is denoted by $\Psuc (\theta) \triangleq \Pr [\SINR_{0} > \theta, \SINR_{\widehat{m}_{0}} > \theta]$ for a given SINR threshold $\theta$; without loss of generality, we consider the same SINR threshold for the two hops.  Let $\Psuc^{(1)} (\theta) \triangleq \Pr [\SINR_{0} > \theta]$ and $\Psuc^{(2)} (\theta) \triangleq \Pr [\SINR_{\widehat{m}_{0}} > \theta]$ denote the success probabilities of the first and second hop, respectively. Using the Fortuin-Kasteleyn-Ginibre (FKG) inequality \cite{Vaz11}, the success probability over the two hops can be bounded as $\Psuc (\theta) \geq \Psuclb (\theta)$, with} \vspace{-1mm}
\begin{align} \label{eq:P_suc}
\red{\Psuclb (\theta) = \Psuc^{(1)} (\theta) \Psuc^{(2)} (\theta)}.
\end{align}
\red{This more tractable expression is obtained by neglecting the spatial correlation between the UL and DL transmissions, i.e., the transmissions in the first and second hop are assumed to occur over two uncorrelated instances of $\Phi_{\mathrm{m}}$ \cite{Atz15a}. Furthermore, if the FD BSs serve two UL/DL nodes with no relaying purposes, the first and second hops become two independent transmissions with success probabilities $\Psuc^{(1)} (\theta)$ and $\Psuc^{(2)} (\theta)$, respectively. We also refer to \cite{Hae13} for the analysis of correlated transmissions in random networks.}


In the rest of the section, we assume that the FD BSs adopt maximum ratio combining (MRC) and maximum ratio transmission (MRT), i.e., the receive combining and transmit beamforming vectors are given by
\begin{align}
\v_{x} = \frac{\h_{\widetilde{m}_{x} x}}{\| \h_{\widetilde{m}_{x} x} \|}, \qquad \w_{x} = \frac{\h_{x \widehat{m}_{x}}}{\| \h_{x \widehat{m}_{x}} \|}
\end{align}
respectively. Different combining configurations are considered in Section~\ref{sec:IC} to study the impact of interference cancellation at the receiver.

\subsection{Success Probability of the First Hop} \label{sec:SP_1}

In this section, we analyze the success probability of the first hop $\Psuc^{(1)} = \Pr [\SINR_{0} > \theta]$, i.e., the probability of successful transmission from the typical HD UL node to the typical FD BS. Considering $\SINR_{0}$ in \eqref{eq:SINR_1}, since MRC is adopted, we have $S_{\widetilde{m}_{0} 0} \sim \chi_{2 N_{\rmR}}^{2}$ (desired signal) and $S_{x 0} \sim \chi_{2}^{2}$, $\forall x \in \Phi \cup \widetilde{\Phi}$ (interferers).\footnote{\label{fn:chi2N}We define a random variable $X \sim \chi_{2 N}^{2}$ to have probability density function (PDF) $f_{X}(x) = \frac{x^{N-1} e^{-x}}{\Gamma(N)}$; its CCDF is given by $\bar{F}_{X}(x) = 1 - \frac{\gamma(N,x)}{\Gamma(N)} = e^{-x} \sum_{n=0}^{N-1} \frac{x^{n}}{n!}$.} Regarding the \red{SI} power $S_{0 0}$, the following lemma provides a tight approximation of the distribution of the \red{SI} power under Rician fading.
\begin{lemma} \label{lem:SI} \rm{
Let $\v_{x}$, $\w_{x}$, and $\H_{x x}$ be independent; in addition, assume that the (non-normalized) elements of $\v_{x}$ and $\w_{x}$ are distributed independently as $\setC \setN (0,1)$. Then, the \red{SI} power $S_{x x} = |\v_{x}^{\herm} \H_{x x} \w_{x}|^{2}$ approximately follows a gamma distribution, i.e., $S_{x x} \sim \Gamma (a,b)$, with shape parameter $a$ and scale parameter $b$ given by\footnote{We define a random variable $X \sim \Gamma(a,b)$ with shape parameter $a$ and scale parameter $b$ to have PDF $f_{X}(x) = \frac{x^{a-1} e^{-x/b}}{b^a \Gamma(a)}$.}
\begin{align} \label{eq:ab}
a \triangleq \frac{(\mu^{2} + \nu^{2})^{2}}{\eta \mu^{4} + 2 \mu^{2} \nu^{2} + \nu^{4}}, \qquad
b \triangleq \frac{\eta \mu^{4} + 2 \mu^{2} \nu^{2} + \nu^{4}}{\mu^{2} + \nu^{2}}
\end{align}
respectively, where $\mu$ and $\nu$ are the \red{absolute mean} and standard deviation, respectively, of the \red{SI} channel $\H_{x x}$ (see \eqref{eq:mu_nu}) and where we have defined
\begin{align}
\eta \triangleq \frac{4 N_{\rmR} N_{\rmT} - (N_{\rmR}+1) (N_{\rmT}+1)}{(N_{\rmR}+1) (N_{\rmT}+1)}.
\end{align}}
\end{lemma}

\begin{IEEEproof}
See Appendix~\ref{sec:A_SI}.
\end{IEEEproof} \vspace{1mm}

\noindent \red{The result of Lemma~\ref{lem:SI} was first derived in our previous work \cite{Atz15a}. Subsequently, it was extended to the case of multi-user MIMO in \cite{Sho17}.}

\begin{remark} \label{rem:SI} \rm{
The assumptions of Lemma~\ref{lem:SI} are very mild. First of all, $\v_{x}$ and $\w_{x}$ are generally chosen as the result of some linear processing of, respectively, $\h_{\widetilde{m}_{x} x}$ and $\h_{x \widehat{m}_{x}}$ (such channels are subject to Rayleigh fading by assumption) followed by power normalization. Note that the assumption of $\v_{x}$ and $\w_{x}$ being MRC and MRT vectors is not required. Besides, $\h_{\widetilde{m}_{x} x}$ and $\h_{x \widehat{m}_{x}}$ are independent on one another and on $\H_{x x}$, and the same holds for $\v_{x}$ and $\w_{x}$. Note that the only practically relevant case where the above assumptions are not satisfied is when $\v_{x}$ (resp. $\w_{x}$) zero-forces the equivalent \red{SI} channel $\H_{x x} \w_{x}$ (resp. $\v_{x}^{\herm} \H_{x x}$): however, this case trivially implies $S_{x x} = 0$ (this scenario is examined in Section~\ref{sec:IC_2}).}
\end{remark}

\begin{figure}[t!]
\centering
\includegraphics[scale=0.9]{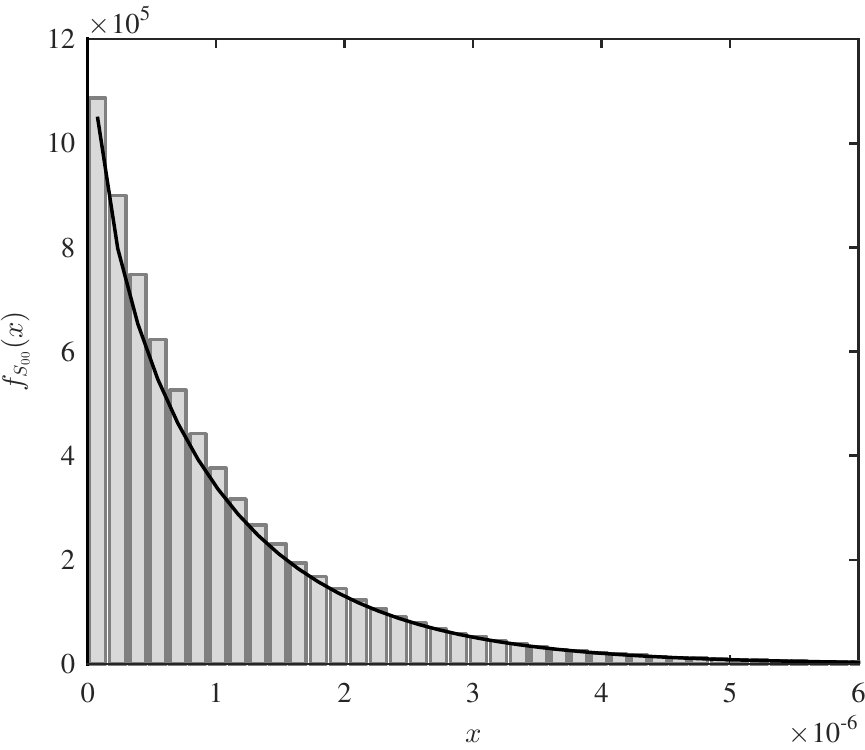}
\caption{PDF of the \red{SI} power for $N_{\rmR} = N_{\rmT} = 4$, $K=1$, and $\Omega = - 60$~dB: empirical histogram versus analytical approximation from Lemma~\ref{lem:SI}.} \label{fig:distr_SI}
\end{figure}

\noindent Lemma~\ref{lem:SI} represents a key result of this paper since it provides a formal characterization of the \red{SI} power experienced by a FD MIMO node with arbitrary receive combining and transmit beamforming vectors. Such distribution of the \red{SI} power is based uniquely on the knowledge of the parameters $K$ and $\Omega$, whose values are available either by design or by measurements, and can be readily incorporated into existing frameworks for the performance analysis of UDNs. Figure~\ref{fig:distr_SI} shows the accuracy of the approximated distribution derived in Lemma~\ref{lem:SI}.
					      
The next theorem provides the success probability of the first hop.

\begin{theorem} \label{th:P_suc1} \rm{
Consider the interference term $I_{0}$ in \eqref{eq:I_1}. The success probability of the first hop is given by
\begin{align} \label{eq:P_suc1}
\Psuc^{(1)} (\theta) & = \sum_{n=0}^{N_{\rmR} - 1} \bigg[ \frac{(- s)^{n}}{n!} \frac{\mathrm{d}^{n}}{\mathrm{d} s^{n}} \mathcal{L}_{I_{0}}(s) \bigg]_{s = \theta \widetilde{\rho}^{-1} \widetilde{R}^{\alpha}}
\end{align}
where
\begin{align} \label{eq:LI_1}
\mathcal{L}_{I_{0}}(s) \triangleq \frac{1}{(1 + s b \widehat{\rho})^{a}} \exp \big( - \lambda \Upsilon(s) \big)
\end{align}
is the Laplace transform of $I_{0}$, where we have defined
\begin{align} \label{eq:Upsilon}
\Upsilon (s) \triangleq 2 \pi \int_{0}^{\infty} \bigg( 1 - \frac{1}{1 + s \widehat{\rho} r^{-\alpha}} \Psi(s, r) \bigg) r \mathrm{d} r
\end{align}
with
\begin{align} \label{eq:Psi}
\Psi (s, r) \triangleq \frac{1}{2 \pi} \int_{0}^{2 \pi} \frac{\mathrm{d} \varphi}{1 + s \widetilde{\rho} (r^{2} + \widetilde{R}^{2} + 2 r \widetilde{R} \cos \varphi)^{- \frac{\alpha}{2}}}.
\end{align}
}
\end{theorem}

\begin{IEEEproof}
See Appendix~\ref{sec:A_P_suc1_th}.
\end{IEEEproof} \vspace{1mm}

\noindent The array gain resulting from the employment of multiple receive antennas appears evident from Theorem~\ref{th:P_suc1}: the larger is $N_{\rmR}$, the more terms are included in the summation of $\Psuc^{(1)}(\theta)$ in \eqref{eq:P_suc1} (the same applies for Theorem~\ref{th:P_suc2}); note that all terms in the summation are positive since the $n$-th derivative of $\mathcal{L}_{I_{0}}(s)$ are negative for odd $n$.

Expressions of the form of \eqref{eq:P_suc1} arise frequently, among other cases, when multiple antennas are involved, and are widely used throughout the paper. A useful upper bound for this type of expression is provided in the following proposition.\footnote{A lower bound with a similar expression can be also obtained; however, such bound is usually not sufficiently tight and it is thus not considered.}

\begin{proposition} \label{pro:Alzer} \rm{
For any $\setL_{X}(s^{\prime}) \triangleq \Exp_{X} \big[ e^{-s^{\prime} X} \big]$ and $N > 1$, the following inequality holds:
\begin{align} \label{eq:Alzer}
\sum_{n=0}^{N - 1} \bigg[ \frac{(- s)^{n}}{n!} \frac{\mathrm{d}^{n}}{\mathrm{d} s^{n}} \mathcal{L}_{X}(s) \bigg]_{s = s^{\prime}} < \sum_{n=1}^{N} (-1)^{n - 1} {{N}\choose{n}} \setL_{X} \big( n \big( \Gamma(N + 1) \big)^{- \frac{1}{N}} s^{\prime} \big).
\end{align}}
\end{proposition}

\begin{IEEEproof}
See Appendix~\ref{sec:A_Alzer}.
\end{IEEEproof} \vspace{1mm}

Given the integral form of $\Upsilon (s)$ in \eqref{eq:Upsilon}, the success probability $\Psuc^{(1)}(\theta)$ is not in closed form and needs to be evaluated numerically; nonetheless, we derive the following closed-form lower and upper bounds.

\begin{corollary} \label{cor:P_suc1} \rm{
The Laplace transform of $I_{0}$ in \eqref{eq:LI_1} is bounded as $\mathcal{L}_{I_{0}}(s) \in \big[ \mathcal{L}_{I_{0}}^{(\min)}(s), \mathcal{L}_{I_{0}}^{(\max)}(s) \big]$, with
\begin{align}
\label{eq:LI_1min} \mathcal{L}_{I_{0}}^{(\min)}(s) & \triangleq \frac{1}{(1 + s b \widehat{\rho})^{a}} \exp \big( - \lambda \Upsilon^{(\max)}(s) \big), \\
\label{eq:LI_1max} \mathcal{L}_{I_{0}}^{(\max)}(s) & \triangleq \frac{1}{(1 + s b \widehat{\rho})^{a}} \exp \big( - \lambda \Upsilon^{(\min)}(s) \big)
\end{align}
where we have defined
\begin{align}
\label{eq:Upsilon_min} \Upsilon^{(\min)}(s) & \triangleq (1 + \tfrac{2}{\alpha}) (\widetilde{\rho}^{\frac{2}{\alpha}} + \widehat{\rho}^{\frac{2}{\alpha}}) \frac{\pi^{2} s^{\frac{2}{\alpha}}}{\alpha \sin \big( \frac{2 \pi}{\alpha} \big)}, \\
\label{eq:Upsilon_max} \Upsilon^{(\max)}(s) & \triangleq 2 (\widetilde{\rho}^{\frac{2}{\alpha}} + \widehat{\rho}^{\frac{2}{\alpha}}) \frac{\pi^{2} s^{\frac{2}{\alpha}}}{\alpha \sin \big( \frac{2 \pi}{\alpha} \big)}.
\end{align}
Then, the lower and upper bounds on the success probability of the first hop $\Psuc^{(1)} (\theta)$, \red{denoted by $\Psuc^{(1,\min)} (\theta)$ and $\Psuc^{(1,\max)} (\theta)$,} are obtained by replacing $\mathcal{L}_{I_{0}}(s)$ in \eqref{eq:P_suc1} with $\mathcal{L}_{I_{0}}^{(\min)}(s)$ and $\mathcal{L}_{I_{0}}^{(\max)}(s)$, respectively.
}
\end{corollary}

\begin{IEEEproof}
See Appendix~\ref{sec:A_P_suc1_cor}.
\end{IEEEproof} \vspace{1mm}

\begin{remark} \rm{
In order to efficiently compute the derivatives of the bounds \eqref{eq:LI_1min}--\eqref{eq:LI_1max}, one can resort to the well-known general Leibniz rule for the differentiation of the product of two functions $f(s) g(s)$ \cite[Eq.~3.3.8]{Abr72}: for instance, for $\mathcal{L}_{I_{0}}^{(\min)}(s)$, we can write $f(s) = \frac{1}{(1 + s b \widehat{\rho})^{a}}$ and $g(s) = \exp \big( -\lambda \Upsilon^{(\max)}(s) \big)$. In turn, the derivatives of $g(s)$ can be computed using Fa\`{a} di Bruno's formula \cite{Joh02} for the differentiation of the composition of two functions $g(s) = (g_{1} \circ g_{2})(s)$, with $g_{1}(s) = \exp(s)$ and $g_{2}(s) = - \lambda \Upsilon^{(\max)}(s)$. These considerations apply equivalently to the bounds provided in Corollary~\ref{cor:P_suc2}.}
\end{remark}

The following corollary provides a sufficient condition under which FD mode outperforms HD mode in terms of spectral efficiency for the case of single receive antenna.

\begin{corollary} \label{cor:HD} \rm{
Consider the first hop assuming that $N_{\rmR}=1$. The achievable spectral efficiency when the BSs operate in FD mode is lower bounded by
\begin{align} \label{eq:T_FD}
\mathrm{SE}_{\mathrm{FD}}^{(\min)}(\theta) \triangleq 2 \mathcal{L}_{I_{0}}^{(\min)}(\theta \widetilde{\rho}^{-1} \widetilde{R}^{\alpha}) \log_{2}(1+\theta).
\end{align}
When the BSs operate in HD mode (i.e., when $\widehat{\rho}=0$), the achievable spectral efficiency is given by
\begin{align} \label{eq:T_HD}
\mathrm{SE}_{\mathrm{HD}}(\theta) \triangleq \exp \bigg( - \lambda \frac{2 \pi^{2} (\theta \widetilde{R}^{\alpha})^{\frac{2}{\alpha}}}{\alpha \sin \big( \frac{2 \pi}{\alpha} \big)} \bigg) \log_{2}(1+\theta).
\end{align}
Then, $\mathrm{SE}_{\mathrm{FD}}^{(\min)}(\theta) \geq \mathrm{SE}_{\mathrm{HD}}(\theta)$ whenever the density $\lambda$ satisfies
\begin{align}
\lambda \leq \frac{\alpha \sin \big( \frac{2 \pi}{\alpha} \big)}{2 \pi^{2} (\theta \widetilde{\rho}^{-1} \widehat{\rho} \widetilde{R}^{\alpha})^{\frac{2}{\alpha}}} \log \bigg( \frac{2}{(1 + b \theta \widetilde{\rho}^{-1} \widehat{\rho} \widetilde{R}^{\alpha})^{a}} \bigg).
\end{align}}
\end{corollary}

\begin{IEEEproof}
The proof is straightforward from Theorem~\ref{th:P_suc1} and Corollary~\ref{cor:P_suc1}.
\end{IEEEproof} \vspace{1mm}

\noindent Evidently, if the density $\lambda$ exceeds a certain threshold, using twice the bandwidth in FD mode does not compensate for the additional interference due to the concurrent UL/DL transmissions and, therefore, HD mode becomes optimal.

\subsection{Success Probability of the Second Hop} \label{sec:SP_2}

In this section, we analyze the success probability of the second hop $\Psuc^{(2)}(\theta) = \Pr [\SINR_{\widehat{m}_{0}} > \theta]$, i.e., the probability of successful transmission from the typical FD BS to the typical HD DL node. Considering $\SINR_{\widehat{m}_{0}}$ in \eqref{eq:SINR_2}, since MRT is adopted, we have $S_{0 \widehat{m}_{0}} \sim \chi_{2 N_{\rmT}}^{2}$ (desired signal) and $S_{x \widehat{m}_{0}} \sim \chi_{2}^{2}$, $\forall x \in \Phi \cup \widetilde{\Phi}$ (interferers). The success probability of the second hop is given next in Theorem~\ref{th:P_suc2}, whereas its lower and upper bounds are provided in Corollary~\ref{cor:P_suc2}.

\begin{theorem} \label{th:P_suc2} \rm{
Consider the interference term $I_{\widehat{m}_{0}}$ in \eqref{eq:I_2}. The success probability of the second hop is given by
\begin{align} \label{eq:P_suc2}
\Psuc^{(2)} (\theta) = \sum_{n=0}^{N_{\rmT} - 1} \bigg[ \frac{(- s)^{n}}{n!} \frac{\mathrm{d}^{n}}{\mathrm{d} s^{n}} \mathcal{L}_{I_{\widehat{m}_{0}}}(s) \bigg]_{s = \theta \widehat{\rho}^{-1} \widehat{R}^{\alpha}}
\end{align}
where
\begin{align} \label{eq:LI_2}
\mathcal{L}_{I_{\widehat{m}_{0}}}(s) \triangleq \Psi(s, \widehat{R}) \exp \big( -\lambda \Upsilon(s) \big)
\end{align}
is the Laplace transform of $I_{\widehat{m}_{0}}$, with $\Upsilon(s)$ and $\Psi (s, r)$ defined in \eqref{eq:Upsilon} and in \eqref{eq:Psi}, respectively.
}
\end{theorem}

\begin{IEEEproof}
See Appendix~\ref{sec:A_P_suc2_th}.
\end{IEEEproof} \vspace{1mm}

\begin{corollary} \label{cor:P_suc2} \rm{
The Laplace transform of $I_{\widehat{m}_{0}}$ in \eqref{eq:LI_2} is bounded as $\mathcal{L}_{I_{\widehat{m}_{0}}}(s) \in \big[ \mathcal{L}_{I_{\widehat{m}_{0}}}^{(\min)}(s), \mathcal{L}_{I_{\widehat{m}_{0}}}^{(\max)}(s) \big]$, with
\begin{align}
\label{eq:LI_2min} \hspace{-1mm} \mathcal{L}_{I_{\widehat{m}_{0}}}^{(\min)}(s) & \triangleq \frac{1}{1 + s \widetilde{\rho} | \widetilde{R} - \widehat{R} |^{-\alpha}} \exp \big( - \lambda \Upsilon^{(\max)}(s) \big), \\
\label{eq:LI_2max} \hspace{-1mm} \mathcal{L}_{I_{\widehat{m}_{0}}}^{(\max)}(s) & \triangleq \frac{1}{1 + s \widetilde{\rho} (\widetilde{R} + \widehat{R})^{-\alpha}} \exp \big( - \lambda \Upsilon^{(\min)}(s) \big)
\end{align}
with $\Upsilon^{(\min)}(s)$ and $\Upsilon^{(\max)}(s)$ defined in \eqref{eq:Upsilon_min} and in \eqref{eq:Upsilon_max}, respectively. Then, the lower and upper bounds on the success probability of the second hop $\Psuc^{(2)} (\theta)$, \red{denoted by $\Psuc^{(2,\min)} (\theta)$ and $\Psuc^{(2,\max)} (\theta)$,} are obtained by replacing $\mathcal{L}_{I_{\widehat{m}_{0}}}(s)$ in \eqref{eq:P_suc2} with $\mathcal{L}_{I_{\widehat{m}_{0}}}^{(\min)}(s)$ and $\mathcal{L}_{I_{\widehat{m}_{0}}}^{(\max)}(s)$, respectively.
}
\end{corollary}

\begin{IEEEproof}
See Appendix~\ref{sec:A_P_suc2_cor}.
\end{IEEEproof} \vspace{1mm}

\noindent Observe that the bounds \eqref{eq:LI_2min} and \eqref{eq:LI_2max} are more accurate when $\widetilde{R} \gg \widehat{R}$ or $\widetilde{R} \ll \widehat{R}$ due to the presence of the first multiplicative term. This condition is easily verified, for instance, when the HD UL node (resp. HD DL node) is a backhaul node and the HD DL node (resp. HD UL node) is a mobile UE, with the former being likely much farther away from the FD small-cell BS with respect to the latter.

\section{Interference Cancellation} \label{sec:IC}

In the previous section, we have considered a MRC/MRT configuration at the FD BSs. This section analyzes interference cancellation at the receive side of both the FD BSs and the HD DL nodes; for the latter, we further extend our analysis to the case of HD DL nodes with multiple receive antennas. We consider PZF, which represents an efficient and low-complexity spatial interference cancellation technique for multi-antenna receivers \cite{Atz16b,Jin08}. If a node is equipped with $N$ receive antennas, the PZF receiver allows to cancel $M \leq N - 1$ interference contributions while using the remaining degrees of freedom to boost the desired received signal.\footnote{\red{Under PZF, the receive combining vector is given by $\v(M) \triangleq (\I_{N} - \Q(M) \Q^{\sharp}(M)) \h_{0} / \| (\I_{N} - \Q(M) \Q^{\sharp}(M)) \h_{0} \|$, where $(\cdot)^{\sharp}$ denotes the Moore-Penrose pseudoinverse operator, $\I_{N}$ is the $N$-dimensional identity matrix, and the columns of $\Q(M) \in \Compl^{N \times M}$ are the effective channels to be cancelled at the receiver.}} Observe that, when $M=0$, the PZF receiver reduces to the MRC case analyzed in Section~\ref{sec:SP_1}. \red{The study of imperfect channel estimation goes beyond the scope of this paper; we refer to \cite{Hua12} for the performance analysis of the PZF receiver with imperfect channel estimation in random networks.}

\red{The PZF receiver requires only the knowledge of the channels to be zero-forced at each time slot: these are generally very few in our setting due to the low-to-moderate number of receive antennas at the small-cell BSs. On the other hand, it is known that the optimal tradeoff between array gain and interference cancellation is achieved by the minimum mean-square error (MMSE) receiver. However, in a fully decentralized network such as that considered in this paper, MMSE receiver may not be practical since it requires the knowledge of the spatial covariance of the interference at each time slot (which depends on the channels and distances of all interfering nodes in the network). In addition, adopting PZF is more relevant for the purpose of our study since it allows us to identify which interference terms are most critical for the deployment of FD technology and their interplay. We refer to \cite{Jin08,Ali10} for the performance analysis of the MMSE receiver in random networks and its comparison with the PZF receiver.}

Focusing on the first hop, we consider two possible receive configurations:
\begin{itemize}
\item[\textit{1)}] Each FD BS cancels the interference coming from the $M$ nearest FD BSs (cf. Section~\ref{sec:IC_1});
\item[\textit{2)}] Each FD BS cancels the \red{SI} (cf. Section~\ref{sec:IC_2}).
\end{itemize}
Observe that the above configurations can be also combined, e.g., by simultaneously cancelling the $M-1$ nearest nodes and the \red{SI}. Focusing on the second hop, we assume multiple receive antennas at the HD DL nodes and consider the following receive configuration:
\begin{itemize}
\item[\textit{3)}] Each HD DL node cancels the inter-node interference (cf. Section~\ref{sec:IC_3}).
\end{itemize}

\subsection{First Hop: Cancelling the Nearest $M$ FD BSs} \label{sec:IC_1}

In interference-limited scenarios, it is often beneficial to cancel the interference coming from a certain number of surrounding nodes. Let us assume that the points of $\Phi$ are indexed such that their distances from the typical FD BS is in increasing order, i.e., $\{ r_{x_{i}} \leq r_{x_{i+1}} \}_{i=1}^{\infty}$, and let us suppose that each FD BS adopts PZF at the receiver to cancel its $M$ nearest FD BSs. The resulting overall interference power at the typical FD BS is given by (cf. \eqref{eq:I_1})
\begin{align} \label{eq:I_1_M}
I_{0}^{\textnormal{\tiny{PZF-$M$}}} & \triangleq \sum_{\substack{x_{i} \in \Phi \\ i > M}} \widehat{\rho} r_{x_{i}}^{-\alpha} S_{x_{i} 0} + \sum_{x_{i} \in \Phi} \widetilde{\rho} r_{\widetilde{m}_{x_{i}}}^{-\alpha} S_{\widetilde{m}_{x_{i}} 0} + \widehat{\rho} S_{0 0}.
\end{align}
The success probability of the first hop with PZF is given in the next theorem.
\begin{theorem} \label{th:P_suc1_M} \rm{
Consider the interference term $I_{0}^{\textnormal{\tiny{PZF-$M$}}}$ in \eqref{eq:I_1_M}. The success probability of the first hop is given by
\begin{align} \label{eq:P_suc1_M}
\Psuc^{(1)}(\theta) = \sum_{n=0}^{N_{\rmR} - M - 1} \bigg[ \frac{(- s)^{n}}{n!} \frac{\mathrm{d}^{n}}{\mathrm{d} s^{n}} \mathcal{L}_{I_{0}^{\textnormal{\tiny{PZF-$M$}}}}(s) \bigg]_{s = \theta \widetilde{\rho}^{-1} \widetilde{R}^{\alpha}}
\end{align}
where
\begin{align} \label{eq:L_I1_M}
\mathcal{L}_{I_{0}^{\textnormal{\tiny{PZF-$M$}}}}(s) \triangleq \frac{1}{(1 + s b \widehat{\rho})^{a}} \Exp_{\Phi} \bigg[ \prod_{\substack{x_{i} \in \Phi \\ i \leq M}} \frac{1}{1 + s \widetilde{\rho} r_{\widetilde{m}_{x_{i}}}^{-\alpha}} \bigg] \Exp_{\Phi} \bigg[ \prod_{\substack{x_{i} \in \Phi \\ i > M}} \frac{1}{1 + s \widehat{\rho} r_{x_{i}}^{-\alpha}} \frac{1}{1 + s \widetilde{\rho} r_{\widetilde{m}_{x_{i}}}^{-\alpha}} \bigg]
\end{align}
is the Laplace transform of $I_{0}^{\textnormal{\tiny{PZF-$M$}}}$.}
\end{theorem}

\begin{IEEEproof}
See Appendix~\ref{sec:A_P_suc1_th_1}.
\end{IEEEproof} \vspace{1mm}

\noindent The tradeoff between array gain and interference cancellation appears evident from Theorem~\ref{th:P_suc1_M}: the larger is $M$, the larger is $\mathcal{L}_{I_{0}^{\textnormal{\tiny{PZF-$M$}}}}(s)$ in \eqref{eq:L_I1_M}, but also the less terms are included in the summation of $\Psuc^{(1)}(\theta)$ in \eqref{eq:P_suc1_M}. Note that a similar expression of the success probability can be obtained if the $M$ nearest HD UL nodes are cancelled.

Unfortunately, a closed-form expression of $\mathcal{L}_{I_{0}^{\textnormal{\tiny{PZF-$M$}}}}(s)$ in \eqref{eq:L_I1_M} is not available and, in order to obtain a more tractable expression, one can resort to the approximation provided in the following corollary.

\begin{corollary} \label{cor:P_suc1_M} \rm{
The Laplace transform of $I_{0}^{\textnormal{\tiny{PZF-$M$}}}$ in \eqref{eq:L_I1_M} can be tightly approximated by
\begin{align} \label{eq:LI_1_M_a}
\mathcal{L}_{I_{0}^{\textnormal{\tiny{PZF-$M$}}}}(s) \simeq \frac{1}{(1 + s b \widehat{\rho})^{a}} \exp \big( - \lambda (\Upsilon_{1}(s,M) + \Upsilon_{2}(s,M)) \big)
\end{align}
where we have defined
\begin{align}
\label{eq:Upsilon1} \Upsilon_{1}(s, M) & \triangleq 2 \pi \int_{0}^{d_{M}} \big( 1 - \Psi (s, r) \big) r \diff r, \\
\label{eq:Upsilon2} \Upsilon_{2}(s, M) & \triangleq 2 \pi \int_{d_{M}}^{\infty} \bigg( 1 - \frac{1}{1 + s \widehat{\rho} r^{-\alpha}} \Psi (s, r) \bigg) r \diff r
\end{align}
with $\Psi (s, r)$ defined in \eqref{eq:Psi} and
\begin{align} \label{eq:d_M}
d_{M} \triangleq (\lambda \pi)^{-\frac{1}{2}} \frac{\Gamma (M + \tfrac{1}{2})}{\Gamma(M)}.
\end{align}}
\end{corollary}

\begin{IEEEproof}
The approximation in \eqref{eq:LI_1_M_a} is obtained using the framework \cite{Atz16b} where $d_{M}$ in \eqref{eq:d_M} is the average distance between the typical FD BS and its $M$-th nearest FD BS, i.e., $d_{M} = \Exp[r_{x_{M}}]$ \cite[Ch.~2.9.1]{Hae12}.
\end{IEEEproof} \vspace{1mm}

\subsection{First Hop: Cancelling the \red{SI}} \label{sec:IC_2}

As discussed in Section~\ref{sec:Intro}, a strong \red{SI} greatly reduces the SINR of the received signals and implicitly sets an upper bound on the transmit power of the FD BSs. Hence, in presence of low \red{SI} attenuation, spatial \red{SI} cancellation at the receiver may be necessary to preserve the SINR of the received signal \red{\cite{Rii11}}. Suppose that each FD BS adopts PZF at the receiver to suppress its \red{SI}.\footnote{Instead of nulling the \red{SI} completely, one can adopt the partial \red{SI} cancellation proposed in \cite{Atz16} at the receiver to enhance the UL throughput.} The resulting overall interference power at the typical FD BS is given by (cf. \eqref{eq:I_1})
\begin{align} \label{eq:I_1_SI}
I_{0}^{\textnormal{\tiny{PZF-SI}}} \triangleq \sum_{x \in \Phi} \big( \widehat{\rho} r_{x}^{-\alpha} S_{x 0} + \widetilde{\rho} r_{\widetilde{m}_{x}}^{-\alpha} S_{\widetilde{m}_{x} 0} \big).
\end{align}
The success probability of the first hop with \red{SI} cancellation is given in the next theorem.
\begin{theorem} \label{th:P_suc1_SI} \rm{
Consider the interference term $I_{0}^{\textnormal{\tiny{PZF-SI}}}$ in \eqref{eq:I_1_SI}. The success probability of the first hop is given by
\begin{align} \label{eq:P_suc1_SI}
\Psuc^{(1)}(\theta) & = \sum_{n=0}^{N_{\rmR} - 2} \bigg[ \frac{(- s)^{n}}{n!} \frac{\mathrm{d}^{n}}{\mathrm{d} s^{n}} \exp \big( - \lambda \Upsilon(s) \big) \bigg]_{s = \theta \widetilde{\rho}^{-1} \widetilde{R}^{\alpha}}.
\end{align}
}
\end{theorem}

\begin{IEEEproof}
The expression in \eqref{eq:P_suc1_SI} can be readily obtained from the proof of Theorem~\ref{th:P_suc1} (see Appendix~\ref{sec:A_P_suc1_th_1}).
\end{IEEEproof} \vspace{1mm}

\noindent Note that removing the \red{SI} greatly simplifies the computation of the success probability.

Now, we wish to answer to the following question: \emph{is it better to use one degree of freedom to suppress the \red{SI} or to cancel the nearest interfering FD BS?} This issue is meaningful in cases where the FD BSs can devote no more than one antenna for interference cancellation (e.g., when the density $\lambda$ is very high). A comparative sufficient condition for this choice is provided in the following corollary.

\begin{corollary} \label{cor:canc_UL} \rm{
Cancelling the \red{SI} is, on average, more beneficial than cancelling the nearest FD BS if
\begin{align} \label{eq:canc_UL}
\frac{4 a}{\pi} \log \bigg( 1 + b \theta \frac{\widehat{\rho}}{\widetilde{\rho}} \widetilde{R}^{\alpha} \bigg) \geq {}_{2}F_{1} \bigg( 1, \frac{2}{\alpha}, 1 + \frac{2}{\alpha}, - \frac{\widetilde{\rho}}{\theta \widehat{\rho} (2 \widetilde{R} \sqrt{\lambda})^{\alpha}} \bigg)
\end{align}
where ${}_{2}F_{1}(a,b,c,x)$ denotes the Gauss hypergeometric function \cite[Sec.~9.1]{Gra07}.}
\end{corollary}

\begin{IEEEproof}
See Appendix~\ref{sec:A_P_suc1_canc1}.
\end{IEEEproof} \vspace{1mm}

\noindent Since the right-hand side of \eqref{eq:canc_UL} is increasing with $\lambda$, Corollary~\ref{cor:canc_UL} formalizes that, on average, the \red{SI} overcomes the interference produced by the nearest FD BS when the density $\lambda$ is below a certain threshold. In fact, the nearest interferer approaches the typical FD BS as the density $\lambda$ increases, and the corresponding average interference power becomes stronger.

\red{Observe that imperfect SI cancellation due to, e.g., imperfect estimation of the SI channel, would still result in a gamma-distributed SI power with coefficients $a$ and $b$ different from those given in Lemma~\ref{lem:SI}. In this respect, building on Lemma~\ref{lem:SI}, the following corollary measures the impact of imperfect estimation of the SI channel on the PZF receiver.}

\begin{corollary} \label{cor:canc_imp_CSI} \rm{
\red{Assume imperfect estimation of the SI channel given by $\hat{\H}_{x x} \triangleq \H_{x x} + \E$, where $\E \in \Compl^{N_{\rmR} \times N_{\rmT}}$ is the estimation error uncorrelated with $\H_{x x}$ and with elements distributed independently as $\setC \setN (0, \epsilon^{2})$. Then, the average SI power after imperfect cancellation of the SI with PZF can be bounded as
\begin{align}
\Exp[S_{x x}] = \Exp \big[ |\v_{x}^{\herm} \H_{x x} \w_{x}|^{2} \big] \leq \epsilon^{2}.
\end{align}}}
\end{corollary}

\begin{IEEEproof}
\red{See Appendix~\ref{sec:A_P_suc1_canc2}.}
\end{IEEEproof} \vspace{1mm}

\subsection{Second Hop: Cancelling the Inter-Node Interference} \label{sec:IC_3}

So far we have assumed single-antenna HD DL nodes. In this section, we extend our analysis to the case of HD DL nodes with multiple receive antennas. First, we aim at answering the following question: \emph{is it better to use one degree of freedom to cancel the inter-node interference or to cancel the nearest interfering FD BS?} A comparative sufficient condition for this choice is provided in the following corollary.

\begin{corollary} \label{cor:canc_DL1} \rm{
Cancelling the inter-node interference is, on average, more beneficial than cancelling the nearest FD BS if
\begin{align} \label{eq:canc_DL1}
\frac{4}{\pi} \log \bigg( 1 + \frac{\widetilde{\rho}}{\widehat{\rho}} \bigg( \frac{\widetilde{R}}{\widehat{R}} + 1 \bigg)^{-\alpha} \bigg) \geq {}_{2}F_{1} \bigg( 1, \frac{2}{\alpha}, 1 + \frac{2}{\alpha}, - \frac{\widehat{\rho}}{\theta \widetilde{\rho} (2 \widehat{R} \sqrt{\lambda})^{\alpha}} \bigg).
\end{align}}
\end{corollary}

\begin{IEEEproof}
See Appendix~\ref{sec:A_P_suc2_canc1}.
\end{IEEEproof} \vspace{1mm}

\noindent The interpretation of Corollary~\ref{cor:canc_DL1} is similar to that of Corollary~\ref{cor:canc_UL} in the sense that, on average, the inter-node interference is stronger than the interference produced by the nearest FD BS when the density $\lambda$ is below a certain threshold.

Of particular interest is the case where the HD UL/DL nodes are mobile UEs served simultaneously within the same small cell. In a small-cell scenario, the distance between UL and DL UEs is generally very short, which causes a severe inter-node interference at the latter \cite{Ale16}. Since the vast majority of commercial mobile UEs is currently equipped with two receive antennas, it is meaningful to examine the impact of inter-node interference cancellation at the HD DL nodes. Therefore, assume that each FD BS serves a pair of HD UL/DL mobile UEs, with the latter equipped with two receive antennas; for simplicity, we suppose that the FD BSs transmit with a single antenna (which encompasses the case of multi-antenna FD BSs performing SDMA). The following corollary provides a sufficient condition for cancelling the inter-node interference, where an evident tradeoff arises between the density $\lambda$ and the radius of the small cell.

\begin{corollary} \label{cor:canc_DL2} \rm{
Suppose that the HD DL node is equipped with two receive antennas and let $\widetilde{R} = \widehat{R} = R$. Cancelling the inter-node interference improves the success probability of the second hop if
\begin{align} \label{eq:canc_DL2}
R \leq \frac{\Delta(\theta)}{\sqrt{\lambda}}
\end{align}
where we have defined
\begin{align}
\Delta(\theta) \triangleq \theta^{1-\frac{1}{\alpha}} \widetilde{\rho} \widehat{\rho}^{\frac{1}{\alpha}-1} \sqrt{\frac{2^{-(2 \alpha + 1)} \alpha}{\Upsilon^{(\min)}(1) (2^{-\alpha} \theta \widetilde{\rho} \widehat{\rho}^{-1} + 1)}}
\end{align}
with $\Upsilon^{(\min)}(s)$ defined in \eqref{eq:Upsilon_min}.}
\end{corollary}

\begin{IEEEproof}
See Appendix~\ref{sec:A_P_suc2_canc2}.
\end{IEEEproof} \vspace{1mm}

\begin{remark} \rm{
Fixing $\widetilde{R} = \widehat{R} = R$ in Corollary~\ref{cor:canc_DL2} models a scenario where the pair of HD UL/DL mobile UEs are located at opposite edges of a small cell with radius $R$. \red{Since the inter-node pathloss is maximized (the distance between HD UL and DL nodes is $2R$), this represents a best-case scenario and inter-node interference cancellation becomes even more desirable for random UE locations within the small cell.} Hence, the condition in \eqref{eq:canc_DL2} can be interpreted as follows: \textit{i)} if the radius $R$ is lower than a certain threshold, then one receive antenna should be invested to cancel the inter-node interference; \textit{ii)} if the density $\lambda$ exceeds a certain threshold, the interference from the other nodes becomes too strong and the two antennas should be used to enhance the signal reception.}
\end{remark}

\section{Numerical Results and Discussion} \label{sec:NUM}

In this section, we present numerical results to assess our theoretical findings. In particular, we aim at answering the following questions: \textit{i)} \textit{under which conditions does FD mode yield performance gains with respect to HD mode?} And \textit{ii)} \textit{what is the impact of interference cancellation at both the FD BSs and the (multi-antenna) DL nodes on the network performance?} These points are addressed next in Sections~\ref{sec:NUM_TG} and \ref{sec:NUM_IC}, respectively. 

Unless otherwise stated, we focus on the scenario where each \red{FD BS} acts as a relay between a HD macro-cell backhaul node and a HD mobile UE. \red{Hence, the macro-cell BSs and the small-cell BSs transmit with powers $\widetilde{\rho} = 43$~dBm and $\widehat{\rho} = 24$~dBm, respectively; the distances of each macro-cell BS and of each mobile UE from their serving small-cell BS are set to $\widetilde{R} = 40$~m and $\widehat{R} = 5$~m, respectively.} \red{We consider the interference-limited case (cf. Section~\ref{sec:SM_SINR}) and set $\sigma^{2}=0$.} The parameters $a$ and $b$ of the \red{SI} are computed according to \eqref{eq:ab} in Lemma~\ref{lem:SI}, where $\mu$ and $\nu$ are obtained from \eqref{eq:mu_nu} with Rician $K$-factor $K=1$ (see \cite{Dua12} for an experimental characterization of $K$) and \red{SI} attenuation $\Omega=-60$~dB. Lastly, the pathloss exponent is $\alpha = 4$ and the SINR threshold is $\theta=0$~dB. 

\begin{figure}[t!]
\centering
\includegraphics[scale=0.9]{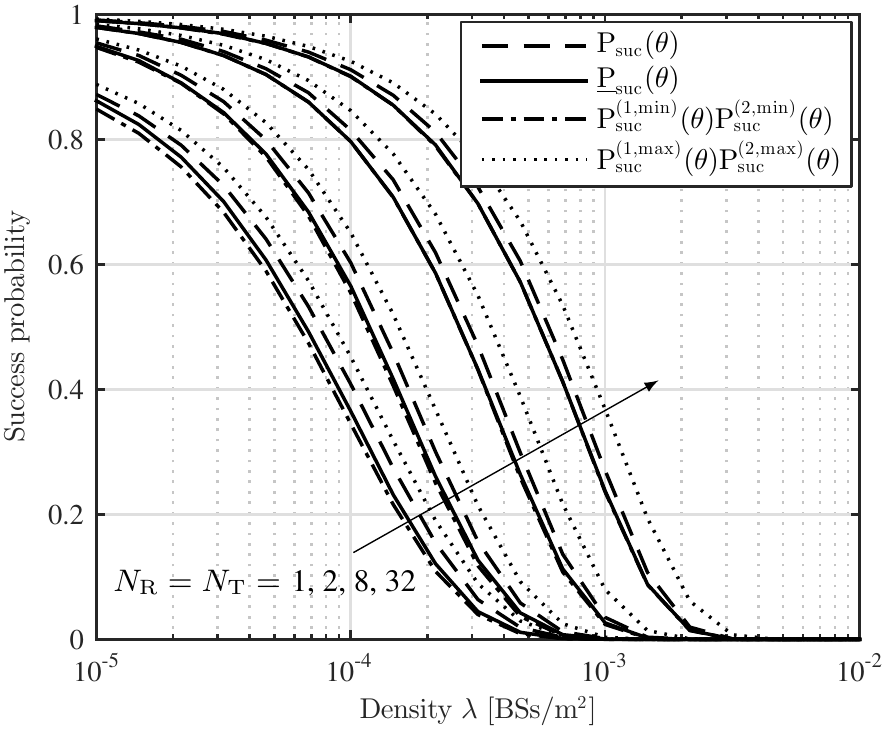}
\caption{Success probability: simulation and analytical bounds versus density $\lambda$, with $\Omega=-60$~dB, $\theta=0$~dB, and for different antenna configurations.} \label{fig:P_suc} \vspace{-3mm}
\end{figure}

\red{We begin by assessing the accuracy of the analytical expressions derived in Section~\ref{sec:SP}. Figure~\ref{fig:P_suc} plots the success probability $\Psuc (\theta)$ and its lower bound $\Psuclb (\theta)$ in \eqref{eq:P_suc}, both obtained by means of Monte Carlo simulations, against the density $\lambda$: these are compared with the lower and upper bounds on $\Psuclb (\theta)$ provided in Corollaries~\ref{cor:P_suc1} and \ref{cor:P_suc2}, respectively. On the one hand, $\Psuclb (\theta)$ is remarkably tight, which justifies our approach of neglecting the spatial correlation between the UL and DL transmissions (cf. Section~\ref{sec:SP}). On the other hand, $\Psuc^{(1,\min)} (\theta) \Psuc^{(2,\min)} (\theta)$ and $\Psuc^{(1,\max)} (\theta) \Psuc^{(2,\max)} (\theta)$ well represent the system performance, also accurately bounding $\Psuc (\theta)$; we also note that the lower bound is increasingly accurate as the number of antennas (at both the receive and the transmit side) increases.} Furthermore, it is evident from Figure~\ref{fig:P_suc} that employing multiple antennas produces substantial SINR gains and compensates for the additional interference generated in FD mode.

\subsection{Throughput Gain} \label{sec:NUM_TG}

We now focus our attention on the first hop in order to analyze the feasibility of FD mode. With this objective in mind, we introduce the \textit{minimum throughput gain} as performance metric, which is defined as
\begin{align}
\mathrm{TG}^{(\min)}(\theta) \triangleq \frac{\mathrm{SE}_{\mathrm{FD}}^{(\min)}(\theta)}{\mathrm{SE}_{\mathrm{HD}}(\theta)}
\end{align}
with $\mathrm{SE}_{\mathrm{FD}}^{(\min)}(\theta)$ and $\mathrm{SE}_{\mathrm{HD}}(\theta)$ defined in \eqref{eq:T_FD} and in \eqref{eq:T_HD}, respectively, for the single-antenna case. This metric represents the worst-case gain of FD mode over HD mode in terms of throughput, with $\mathrm{TG}^{(\min)}(\theta) > 1$ indicating that FD mode outperforms the equivalent HD setup.

\begin{figure}[t!]
\begin{minipage}[c]{0.49\textwidth}
\centering
\includegraphics[scale=0.9]{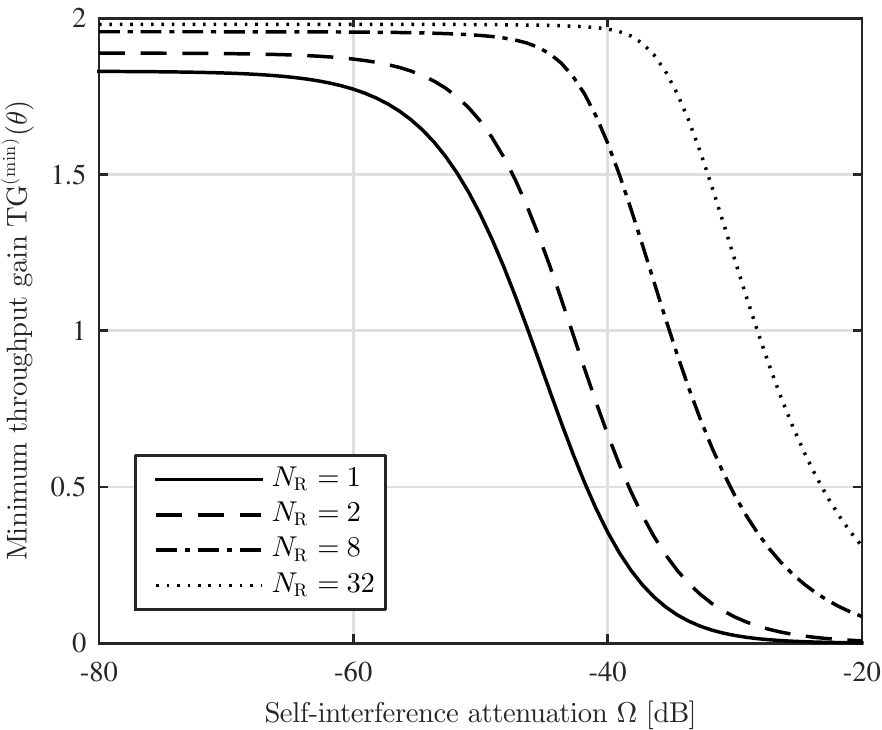} 
\caption{Minimum throughput gain of FD mode over HD mode in the first hop against the \red{SI} cancellation $\Omega$, with $\lambda=10^{-4}$~BSs/m$^{2}$, $\theta=0$~dB, and for different numbers of receive antennas.} \label{fig:TG_Omega} \vspace{-3mm}
\end{minipage}
\hspace{1mm}
\begin{minipage}[c]{0.49\textwidth}
\centering
\includegraphics[scale=0.9]{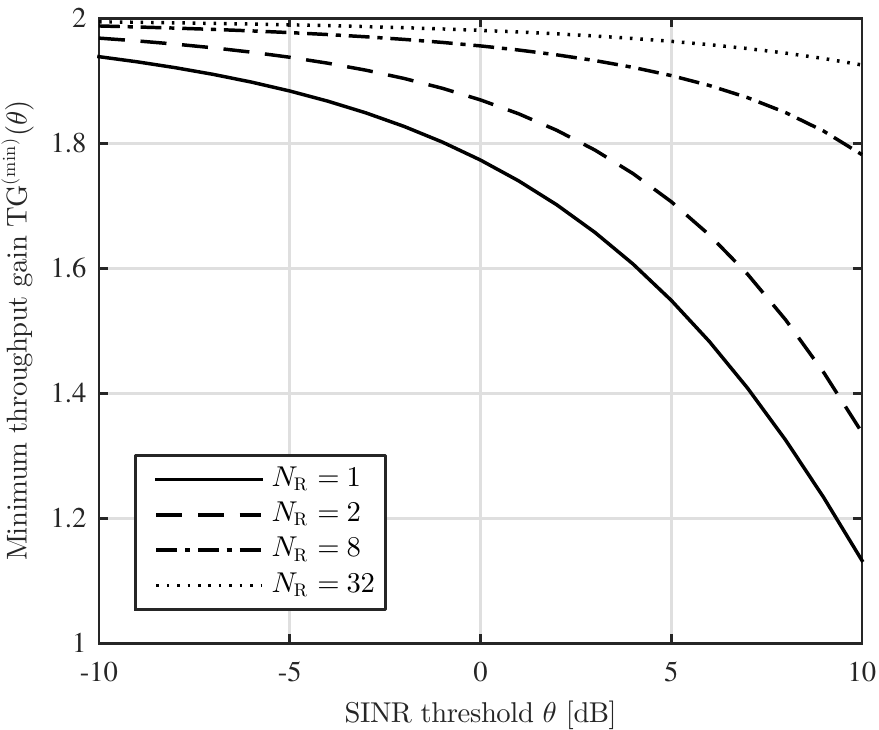} 
\caption{Minimum throughput gain of FD mode over HD mode in the first hop~against the SINR threshold $\theta$, with $\lambda=10^{-4}$~BSs/m$^{2}$, $\Omega=-60$~dB, and for different numbers of receive antennas.} \label{fig:TG_theta} \vspace{-3mm}
\end{minipage}
\end{figure}

Notably, the analytical tools presented in Section~\ref{sec:SP_1} allow to evaluate the effect of multiple receive antennas in mitigating the \red{SI}, which represents a crucial issue in FD communications. Figure~\ref{fig:TG_Omega} plots the minimum throughput gain against the \red{SI} attenuation $\Omega$ with $\lambda=10^{-4}$ and $\theta=0$~dB. Hence, we have $\mathrm{TG}^{(\min)}(\theta) > 1$ even for moderate values of the \red{SI} attenuation, namely: $\Omega \leq -47$~dB for $N_{\rmR}=N_{\rmT}=1$, $\Omega \leq -43$~dB for $N_{\rmR}=N_{\rmT}=2$, $\Omega \leq -35$~dB for $N_{\rmR}=N_{\rmT}=8$, and $\Omega \leq -29$~dB for $N_{\rmR}=N_{\rmT}=32$~dB. On the other hand, the minimum throughput gain is analyzed in Figure~\ref{fig:TG_theta} as a function of the SINR threshold $\theta$ with $\lambda=10^{-4}$ and $\Omega=-60$~dB: in this respect, it is shown that FD mode improves the performance with respect to HD mode for any reasonable value of $\theta$.

\subsection{Interference Cancellation} \label{sec:NUM_IC}

Lastly, we consider interference cancellation at both the FD BSs and at the (multi-antenna) HD DL nodes and analyze its impact on the network performance. In doing so, we make use of the tools developed for PZF receivers proposed in Section~\ref{sec:IC}.

\begin{figure}[t!]
\begin{minipage}[c]{0.49\textwidth}
\centering
\vspace{-3mm}
\includegraphics[scale=0.9]{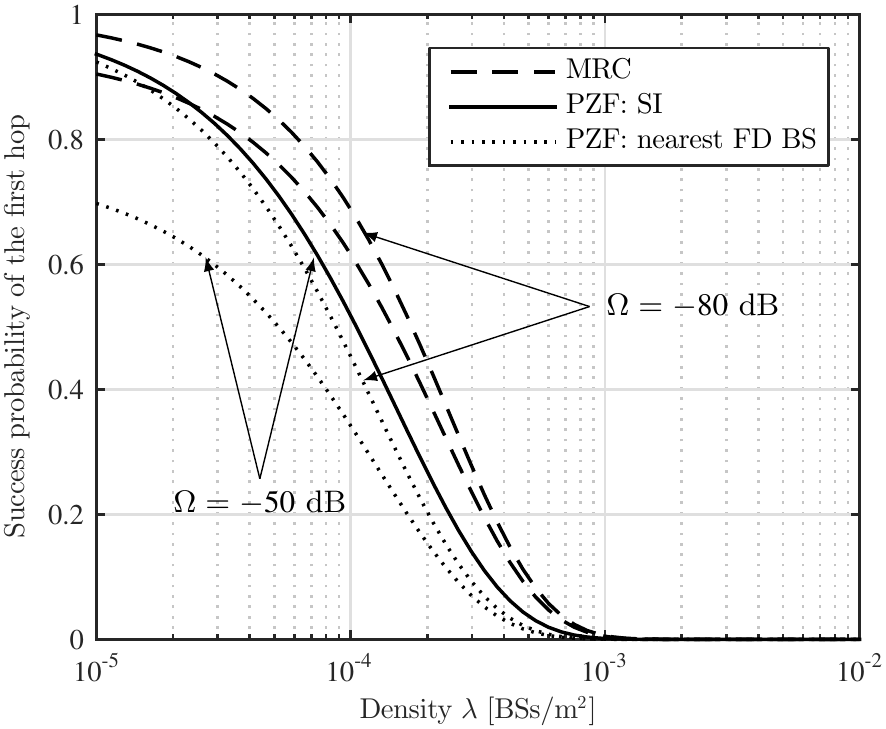} 
\caption{Success probability of the first hop with interference cancellation against the density $\lambda$, with $N_{\rmR}=2$, $\theta=0$~dB, and for two different values of the \red{SI} cancellation $\Omega$.} \label{fig:canc_M} \vspace{-3mm}
\end{minipage}
\hspace{1mm}
\begin{minipage}[c]{0.49\textwidth}
\centering
\includegraphics[scale=0.9]{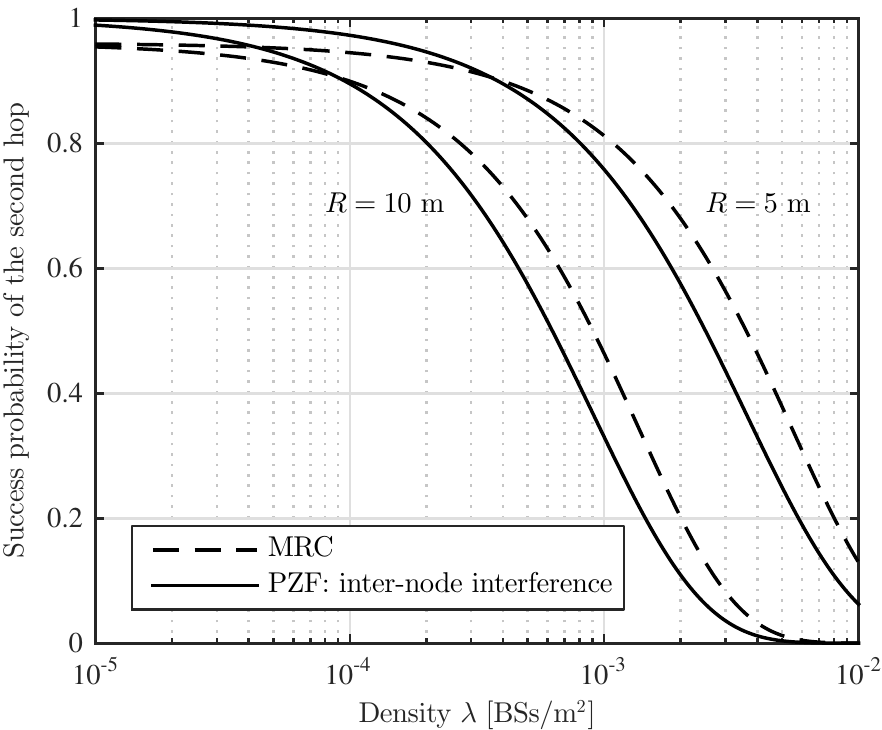} 
\caption{Success probability of the second hop considering DL nodes with two receive antennas against the density $\lambda$, with $\theta=0$~dB and for two different values of the radius of the small cell $R$.} \label{fig:canc_DL} \vspace{-3mm}
\end{minipage}
\end{figure}

Focusing on the first hop, assume that the FD BSs are equipped with $N_{\rmR} = 2$ receive antennas and that they employ PZF to cancel either the \red{SI} or the nearest FD BS. Figure~\ref{fig:canc_M} plots the success probability of the first hop against the density $\lambda$ and compares the above scenarios with the case of no interference cancellation. On the one hand, when the \red{SI} attenuation is low (i.e., $\Omega=-50$~dB), \red{suppressing the \red{SI} improves the performance for densities lower than $\lambda = 2 \times 10^{-3}$~BSs/m$^{2}$,} whereas MRC is the best option for higher densities; on the other hand, for a higher \red{SI} attenuation (i.e., $\Omega=-80$~dB), it is always better to use both antennas for array gain. Hence, since the array gain obtained with just two antennas is significant with respect to the single-antenna case (see Figure~\ref{fig:P_suc}), one should always exploit both antennas for boosting the desired received signal unless $\Omega$ is very low.

Consider now the scenario described in Section~\ref{sec:IC_3}, where each single-antenna FD BS serves a pair of HD UL/DL mobile UEs located at opposite edges of a small cell with radius $R$, with the DL UE equipped with two receive antennas. Figure~\ref{fig:canc_DL} plots the success probability of the second hop against the density $\lambda$ with inter-node interference cancellation using PZF and compares it with the MRC case. Note that the crossing point between the corresponding curves can be recovered exactly from Corollary~\ref{cor:canc_DL2}. As expected, suppressing the inter-node interference becomes detrimental at high densities, where both antennas at the mobile UE should be used for boosting the desired received signal. Recall that this represents a best-case scenario and inter-node interference cancellation becomes even more desirable for random UE locations within the small cell (see Remark~\ref{rem:SI}).

\section{Conclusions} \label{sec:END}

In this paper, we investigate the success probability and spectral efficiency performance of full-duplex (FD) multiple-input multiple-output (MIMO) small-cell networks using tools from stochastic geometry. \red{The proposed framework provides insights into the system-level gains of FD mode with respect to half-duplex mode in terms of network throughput.} In particular, the use of the extra degrees of freedom -- brought by multiple antennas -- for either desired signal power increase or interference cancellation is studied. Simulation results show the beneficial effect of multiple antennas in mitigating the additional interference introduced by FD mode and demonstrate the feasibility of FD technology in practical scenarios even for moderate values of the \red{SI} attenuation. In this respect, partial zero forcing is shown to be a promising antenna processing technique for beneficial FD operation.

\red{Further extensions to this work may include studying the feasibility of FD multi-user MIMO and massive-MIMO systems. It would also be of interest to explore how non-linear interference cancellation techniques and user selection affect the network performance.}

\appendices

\section{Success Probability of the First Hop}
\subsection{Proof of Lemma~\ref{lem:SI}} \label{sec:A_SI}

In this appendix, we derive the approximate distribution of the \red{SI} power $S_{x x}$; for notational simplicity, in the following we omit the sub-indices in the beamforming vectors and in the channel matrix and write $S_{x x} \triangleq |\v^{\herm} \H \w|^{2}$. Let $\v \triangleq (v_{i})_{i=1}^{N_{\rmR}}$, $\w \triangleq (w_{j})_{j=1}^{N_{\rmT}}$, and $\H \triangleq \big( (h_{i j})_{i=1}^{N_{\rmR}} \big)_{j=1}^{N_{\rmT}}$. Hence, assuming that $\v$, $\w$, and $\H$ are independent (see Remark~\ref{rem:SI}),  we can write \vspace{-3mm}
\begin{align} \label{eq:vHw}
|\v^{\herm} \H \w|^{2} = \sum_{i,k=1}^{N_{\rmR}} \sum_{j,\ell=1}^{N_{\rmT}} v_{i}^{*} v_{k} h_{i j} h_{k \ell}^{*} w_{j} w_{\ell}^{*}.
\end{align}
Then, building on the central limit theorem for causal functions \cite{Pap62}, we can approximate a sum of positive random variables $X = \sum_{i} X_{i}$ using the gamma distribution with shape and scale parameters given by \vspace{-3mm}
\begin{align} \label{eq:gamma_ab}
a = \frac{(\Exp[X])^{2}}{\Var[X]}, \qquad b = \frac{\Var[X]}{\Exp[X]}
\end{align}
respectively, with $\Var[X]$ denoting the variance of the random variable $X$.

First of all, we outline the statistics of $\H$, $\v$, and $\w$. Recalling that \red{$h_{i j} \sim \setC \setN (\mu_{i j}, \nu^{2})$, with $| \mu_{i j} | = \mu$, $\forall i=1, \ldots, N_{\rmR}$, $\forall j=1, \ldots, N_{\rmT}$}, we have $\Exp \big[ |h_{i j}|^{2} \big] = \mu^{2} + \nu^{2}$ and $\Exp \big[ |h_{i j}|^{4} \big] = \mu^{4} + 4 \mu^{2} \nu^{2} + 2 \nu^{4}$. On the other hand, since $\| \v \|^{2} = \| \w \|^{2} = 1$, we have $\Exp \big[ |v_{i}|^{2} \big] = \frac{1}{N_{\rmR}}$ and $\Exp \big[ |w_{j}|^{2} \big] = \frac{1}{N_{\rmT}}$. Furthermore, we can write $v_{i} = \frac{\bar{v}_{i}}{\| \bar{\v} \|}$ and $w_{j} = \frac{\bar{w}_{j}}{\| \bar{\w} \|}$, where $\bar{v}_{i}, \bar{w}_{j} \sim \setC \setN (0,1)$ are the non-normalized coefficients of $\v$ and $\w$ (see Remark~\ref{rem:SI}), respectively, with $\bar{\v} \triangleq (\bar{v}_{i})_{i=1}^{N_{\rmR}}$ and $\bar{\w} \triangleq (\bar{w}_{j})_{j=1}^{N_{\rmT}}$. It follows that $|v_{i}|^{4} = \frac{|\bar{v}_{i}|^{4}}{\| \bar{\v} \|^{4}}$, where
\begin{align}
\| \bar{\v} \|^{4} & = (\bar{\v}^{\herm} \bar{\v})^{2} = \sum_{i=1}^{N_{\rmR}} |\bar{v}_{i}|^{4} + \sum_{\substack{i,j=1 \\ i \neq j}}^{N_{\rmR}} |\bar{v}_{i}|^{2} |\bar{v}_{j}|^{2}, \\
\Exp \big[ \| \bar{\v} \|^{4} \big] & = N_{\rmR} \Exp \big[ |\bar{v}_{i}|^{4} \big] + N_{\rmR} (N_{\rmR} - 1) \big( \Exp \big[ |\bar{v}_{i}|^{2} \big] \big)^{2} = N_{\rmR} (N_{\rmR} + 1), \\
\Exp \big[ |v_{i}|^{4} \big] & = \frac{\Exp \big[ |\bar{v}_{i}|^{4} \big]}{N_{\rmR} (N_{\rmR} + 1)} = \frac{2}{N_{\rmR} (N_{\rmR} + 1)}, \\
\Exp \big[ |v_{i}|^{2} |v_{j}|^{2} \big] & = \frac{\big( \Exp \big[ |\bar{v}_{i}|^{2} \big] \big)^{2}}{N_{\rmR} (N_{\rmR} + 1)} = \frac{1}{N_{\rmR} (N_{\rmR} + 1)}
\end{align}
and, likewise, from $|w_{j}|^{4} = \frac{|\bar{w}_{j}|^{4}}{\| \bar{\w} \|^{4}}$, we have $\Exp \big[ |w_{j}|^{4} \big] = \frac{2}{N_{\rmT} (N_{\rmT} + 1)}$ and $\Exp \big[ |w_{i}|^{2} |w_{j}|^{2} \big] = \frac{1}{N_{\rmT} (N_{\rmT} + 1)}$.

In order to obtain the parameters of the gamma function introduced in \eqref{eq:gamma_ab}, we need to derive the second and fourth moments of $|\v^{\herm} \H \w|$. Let us define $\Sigmab \triangleq \diag (\sigma_{i})_{i=1}^{N_{\min}}$, where $\sigma_{i}$ denotes the $i$-th singular value of $\H$ and $N_{\min} \triangleq \min(N_{\rmR}, N_{\rmT})$: recalling \eqref{eq:vHw} and the above properties of the beamforming vectors, we obtain
\begin{align}
\label{eq:vHw1} \Exp \big[ |\v^{\herm} \H \w|^{2} \big] & = \Exp \bigg[ \sum_{i=1}^{N_{\rmR}} \sum_{j=1}^{N_{\rmT}} |v_{i}|^{2} |h_{i j}|^{2} |w_{j}|^{2} \bigg] = \frac{1}{N_{\rmR} N_{\rmT}} \sum_{i=1}^{N_{\rmR}} \sum_{j=1}^{N_{\rmT}} \Exp \big[ |h_{i j}|^{2} \big] = \mu^{2} + \nu^{2}
\end{align}
and
\begin{align}
\nonumber \Exp \big[ |\v^{\herm} \H \w|^{4} \big] & = \Exp \bigg[ \sum_{i=1}^{N_{\min}} |v_{i}|^{4} |\sigma_{i}|^{4} |w_{j}|^{4} \bigg] + 2 \Exp \bigg[ \sum_{\substack{i,j=1 \\ i \neq j}}^{N_{\min}} |v_{i}|^{2} |v_{j}|^{2} |\sigma_{i}|^{2} |\sigma_{j}|^{2} |w_{i}|^{2} |w_{j}|^{2} \bigg] \\
\nonumber & = \frac{2}{N_{\rmR} {N_{\rmT} (N_{\rmR} + 1) (N_{\rmT} + 1)}} \bigg( 2 \Exp \bigg[ \sum_{i=1}^{N_{\min}} |\sigma_{i}|^{4} \bigg] + \Exp \bigg[ \sum_{\substack{i,j=1 \\ i \neq j}}^{N_{\min}} |\sigma_{i}|^{2} |\sigma_{j}|^{2} \bigg] \bigg) \\
\nonumber & = \frac{2}{N_{\rmR} {N_{\rmT} (N_{\rmR} + 1) (N_{\rmT} + 1)}} \bigg( \Exp \bigg[ \sum_{i=1}^{N_{\min}} |\sigma_{i}|^{4} \bigg] + \Exp \bigg[ \sum_{i,j=1}^{N_{\min}} |\sigma_{i}|^{2} |\sigma_{j}|^{2} \bigg] \bigg) \\
\nonumber & = \frac{2}{N_{\rmR} {N_{\rmT} (N_{\rmR} + 1) (N_{\rmT} + 1)}} \Big( \Exp \big[ \tr \big( (\H^{\herm} \H)^{2} \big) \big] + \Exp \big[ \big( \tr (\H^{\herm} \H) \big)^{2} \big] \Big) \\
\label{eq:vHw4} & = \frac{4 N_{\rmR} N_{\rmT}}{(N_{\rmR} + 1) (N_{\rmT} + 1)} \mu^{4} + 4 \mu^{2} \nu^{2} + 2 \nu^{4}
\end{align}
\red{where the second last line in \eqref{eq:vHw4} follows from} the fact that $\Exp \big[ \sum_{i,j} |\sigma_{i}|^{2} |\sigma_{j}|^{2} \big] = \Exp \big[ \sum_{i} |\sigma_{i}|^{2} \big] \Exp \big[ \sum_{j} |\sigma_{j}|^{2} \big]$ and
\begin{align}
\sum_{i=1}^{N_{\min}} |\sigma_{i}|^{2} = \tr (\Sigmab^{\herm} \Sigmab) = \tr (\H^{\herm} \H), \qquad
\sum_{i=1}^{N_{\min}} |\sigma_{i}|^{4} = \tr \big( (\Sigmab^{\herm} \Sigmab)^{2} \big) = \tr \big( (\H^{\herm} \H)^{2} \big)
\end{align}
\red{whereas the last line in \eqref{eq:vHw4} results from}
\begin{align}
\vspace{1mm} \Exp \big[ \big( \tr (\H^{\herm} \H) \big)^2 \big] & \! = \! N_{\rmR} N_{\rmT} \Big( \Exp \big[ |h_{i j}|^{4} \big] \! + \! (N_{\rmR} N_{\rmT} - 1) \Exp \big[ |h_{i j}|^{2} \big] \Big), \\
\vspace{1mm} \Exp \big[ \tr \big( (\H^{\herm} \H)^{2} \big) \big] & \! = \! N_{\rmR} N_{\rmT} \Big( \Exp \big[ |h_{i j}|^{4} \big] \! + \! (N_{\rmR} + N_{\rmT} + 2) \big( \Exp \big[ |h_{i j}|^{2} \big] \big)^{2} \! \! + \! (N_{\rmR} - 1)(N_{\rmT} - 1) \big( \Exp [h_{i j}] \big)^{4} \Big).
\end{align}
Since $\Var \big[ |\v^{\herm} \H \w|^{2} \big] = \Exp \big[ |\v^{\herm} \H \w|^{4} \big] - \big( \Exp \big[ |\v^{\herm} \H \w|^{2} \big] \big)^{2}$, we readily obtain $a$ and $b$ in \eqref{eq:ab} by applying \eqref{eq:gamma_ab}. This concludes the proof. \hfill \IEEEQED

\subsection{Proof of Theorem~\rm{\ref{th:P_suc1}}} \label{sec:A_P_suc1_th}

The success probability of the first hop is given by
\begin{align}
\Psuc^{(1)} (\theta) = \Pr \bigg[ \frac{\widetilde{\rho} \widetilde{R}^{-\alpha} S_{\widetilde{m}_{0} 0}}{I_{0}} > \theta \bigg] = \Pr \big[ S_{\widetilde{m}_{0} 0} > \theta \widetilde{\rho}^{-1} \widetilde{R}^{\alpha} I_{0} \big] = \Exp_{I_{0}} \big[ \bar{F}_{S_{\widetilde{m}_{0} 0}} \big( \theta \widetilde{\rho}^{-1} \widetilde{R}^{\alpha} I_{0} \big) \big]
\end{align}
where $I_{0}$ is defined in \eqref{eq:I_1} and denotes the overall interference at the typical FD BS. Since the latter is equipped with $N_{\rmR}$ receive antennas, the power of its desired signal is distributed as $S_{\widetilde{m}_{0} 0} \sim \chi_{2 N_{\rmR}}^{2}$: hence, our case falls into the general framework \cite{Hun08} and $\Psuc^{(1)}(\theta)$ in \eqref{eq:P_suc1} results from applying (see footnote \ref{fn:chi2N})
\begin{align}
\Exp_{I_{0}} \big[ \bar{F}_{S_{\widetilde{m}_{0} 0}} \big( s I_{0} \big) \big] = \Exp_{I_{0}} \bigg[ e^{- s I_{0}} \sum_{n=0}^{N_{\mathrm{R}}-1} \frac{\big( s I_{0} \big)^{n}}{n!} \bigg] = \sum_{n=0}^{N_{\mathrm{R}}-1} \bigg[ \frac{(-s)^{n}}{n!} \frac{\diff^{n}}{\diff s^{n}} \setL_{I_{0}}(s) \bigg].
\end{align}

On the other hand, building on \cite[Th.~1]{Ton15}, the Laplace transform of $I_{0}$ is obtained as
\begin{align}
\nonumber \mathcal{L}_{I_{0}}(s) & = \Exp \big[ e^{-s I_{0}} \big] \\
\nonumber & = \Exp \big[ \exp (- s \widehat{\rho} S_{0 0}) \big] \Exp \bigg[ \prod_{x \in \Phi} \exp \big( - s (\widehat{\rho} r_{x}^{-\alpha} S_{x 0} + \widetilde{\rho} r_{\widetilde{m}_{x}}^{-\alpha} S_{\widetilde{m}_{x} 0}) \big) \bigg] \\
\nonumber & = \frac{1}{(1 + s b \widehat{\rho})^{a}} \Exp_{\Phi} \bigg[ \prod_{x \in \Phi} \frac{1}{1 + s \widehat{\rho} r_{x}^{-\alpha}} \frac{1}{1 + s \widetilde{\rho} r_{\widetilde{m}_{x}}^{-\alpha}} \bigg] \\
\label{eq:LI_1a} & = \frac{1}{(1 + s b \widehat{\rho})^{a}} \exp \bigg( - \lambda \int_{\Real^{2}} \bigg( 1 - \frac{1}{1 + s \widehat{\rho} r_{x}^{-\alpha}} \frac{1}{1 + s \widetilde{\rho} r_{\widetilde{m}_{x}}^{-\alpha}} \bigg) \diff x \bigg)
\end{align}
\red{where in the second last line in \eqref{eq:LI_1a} we have applied the moment-generating function of the gamma and of the exponential distributions and in the last line in \eqref{eq:LI_1a} we have applied the probability generating functional of a PPP.} Finally, the expression in \eqref{eq:LI_1} follows from
\begin{align}
\label{eq:m_tilde1} r_{\widetilde{m}_{x}} =  \| x + \widetilde{R} (\cos \varphi, \sin \varphi) \| = \sqrt{r_{x}^{2} + \widetilde{R}^{2} + 2 r_{x} \widetilde{R} \cos \varphi}
\end{align}
with $\varphi$ uniformly distributed in $[0, 2\pi]$. This completes the proof. \hfill \IEEEQED

\subsection{Proof of Proposition~\rm{\ref{pro:Alzer}}} \label{sec:A_Alzer}

For any random variable $X$ and $N > 1$, we have that (see footnote~\ref{fn:chi2N})
\begin{align} \label{eq:Alzer1}
\sum_{n=0}^{N - 1} \bigg[ \frac{(- s)^{n}}{n!} \frac{\mathrm{d}^{n}}{\mathrm{d} s^{n}} \mathcal{L}_{X}(s) \bigg]_{s = s^{\prime}} = 1 - \frac{\Exp_{X} \big[ \gamma(N,s^{\prime} X) \big]}{\Gamma(N)}
\end{align}
Now, we use Alzer's inequality \cite{Alz97}, by which
\begin{align}
\frac{\gamma(N,x)}{\Gamma(N)} > (1- e^{-c x})^{N}
\end{align}
with $c \triangleq \big( \Gamma(N+1) \big)^{-\frac{1}{N}}$ for $N > 1$. Now, expanding the expectation term in \eqref{eq:Alzer1}, we have
\begin{align}
\nonumber \frac{\Exp_{X} \big[ \gamma(N, s^{\prime} X) \big]}{\Gamma(N)} & > \Exp_{X} \big[ (1 - e^{- c s^{\prime} X})^{N} \big] \\
\nonumber & = \Exp_{X} \bigg[ \sum_{n=0}^{N} (-1)^{n} \binom{N}{n} e^{- n c s^{\prime} X} \bigg] \\
\label{eq:Alzer2} & = \sum_{n=0}^{N} (-1)^{n} \binom{N}{n} \setL_{X}(n c s^{\prime}).
\end{align}
\red{Finally, the upper bound in \eqref{eq:Alzer} results from plugging the last line in \eqref{eq:Alzer2} into \eqref{eq:Alzer1}}, where $-(-1)^{n} = (-1)^{n-1}$ and $\binom{N}{0} \setL_{X}(0)=1$. \hfill \IEEEQED

\subsection{Proof of Corollary~\rm{\ref{cor:P_suc1}}} \label{sec:A_P_suc1_cor}

Building on \cite[Th.~3]{Ton15}, we have that $\Upsilon(s)$ in \eqref{eq:Upsilon} is bounded as $\Upsilon(s) \in \big[ \Upsilon^{(\min)}(s), \Upsilon^{(\max)}(s) \big]$, with $\Upsilon^{(\min)}(s)$ and $\Upsilon^{(\max)}(s)$ defined in \eqref{eq:Upsilon_min} and in \eqref{eq:Upsilon_max}, respectively. Then, the lower and upper bounds on $\mathcal{L}_{I_{0}}(s)$ in \eqref{eq:LI_1min}--\eqref{eq:LI_1max} readily follow. \hfill \IEEEQED

\section{Success Probability of the Second Hop}
\subsection{Proof of Theorem~\rm{\ref{th:P_suc2}}} \label{sec:A_P_suc2_th}

The success probability of the second hop is given by
\begin{align}
\Psuc^{(2)} (\theta) = \Pr \bigg[ \frac{\widehat{\rho} \widehat{R}^{-\alpha} S_{0 \widehat{m}_{0}}}{I_{\widehat{m}_{0}}} > \theta \bigg] = \Pr \big[ S_{0 \widehat{m}_{0}} > \theta \widehat{\rho}^{-1} \widehat{R}^{\alpha} I_{\widehat{m}_{0}} \big] = \Exp_{I_{\widehat{m}_{0}}} \big[ \bar{F}_{S_{0 \widehat{m}_{0}}} \big( \theta \widehat{\rho}^{-1} \widehat{R}^{\alpha} I_{\widehat{m}_{0}} \big) \big]
\end{align}
where $I_{\widehat{m}_{0}}$ is defined in \eqref{eq:I_2} and denotes the overall interference at the typical HD DL node. Since the typical FD BS is equipped with $N_{\rmT}$ transmit antennas, the power of the desired signal is distributed as $S_{0 \widehat{m}_{o}} \sim \chi_{2 N_{\rmT}}^{2}$ and the expression in \eqref{eq:P_suc2} is obtained following similar steps as in Appendix~\ref{sec:A_P_suc1_th}.

On the other hand, building again on \cite[Th.~1]{Ton15}, the Laplace transform of $I_{\widehat{m}_{0}}$ is obtained as
\begin{align}
\nonumber \mathcal{L}_{I_{\widehat{m}_{0}}}(s) & = \Exp \big[ e^{-s I_{\widehat{m}_{0}}} \big] \\
\label{eq:LI_2a} & = \Exp \big[ \exp (- s \widetilde{\rho} r_{\widetilde{m}_{0}}^{-\alpha} S_{\widetilde{m}_{0} \widehat{m}_{0}}) \big] \Exp \bigg[ \prod_{x \in \Phi} \exp \big( - s (\widehat{\rho} r_{x}^{-\alpha} S_{x \widehat{m}_{0}} + \widetilde{\rho} r_{\widetilde{m}_{x}}^{-\alpha} S_{\widetilde{m}_{x} \widehat{m}_{0}}) \big) \bigg]
\end{align}
where $\Exp \big[ \exp (- s \widetilde{\rho} r_{\widetilde{m}_{0}}^{-\alpha} S_{\widetilde{m}_{0} \widehat{m}_{0}}) \big] = \Psi(s,\widehat{R})$ from the moment-generating function of the exponential distribution and \eqref{eq:m_tilde1}, \red{and where the second expectation term is equivalent to that in the second line in \eqref{eq:LI_1a}}. This concludes the proof. \hfill \IEEEQED

\subsection{Proof of Corollary~\rm{\ref{cor:P_suc2}}} \label{sec:A_P_suc2_cor}

Given the definition of $\Psi(s,r)$ in \eqref{eq:Psi}, it is not difficult to find the following bounds:
\begin{align} \label{eq:Psi_bounds}
\Psi(s,r) \in \bigg[ \frac{1}{1 + s \widetilde{\rho} |r - \widetilde{R}|^{-\alpha}} , \frac{1}{1 + s \widetilde{\rho} (r + \widetilde{R})^{-\alpha}} \bigg].
\end{align}
Then, the lower and upper bounds on $\mathcal{L}_{I_{\widehat{m}_{0}}}(s)$ in \eqref{eq:LI_2min}--\eqref{eq:LI_2max} are a straightforward result of combining \eqref{eq:Psi_bounds} and Corollary~\ref{cor:P_suc1}. \hfill \IEEEQED

\section{Interference Cancellation}
\subsection{Proof of Theorem~\ref{th:P_suc1_M}} \label{sec:A_P_suc1_th_1}

The proof follows similar steps as in Appendix~\ref{sec:A_P_suc1_th}. By applying PZF, the typical FD BS uses $N_{\rmR}-M$ receive antennas to match its desired received signal and, therefore, the power of the latter is distributed as $S_{\widetilde{m}_{0} 0} \sim \chi_{2 (N_{\rmR}-M)}^{2}$, yielding the expression in \eqref{eq:P_suc1_M} (see \cite{Atz16b} for details). On the other hand, the Laplace transform of $I_{0}^{\textnormal{\tiny{PZF-$M$}}}$ in \eqref{eq:L_I1_M} is obtained by removing the interference contribution of the first $M$ FD BSs and \red{the expectation term in the second last line in \eqref{eq:LI_1a} becomes}
\begin{align} \label{eq:L_I1_M1}
\Exp_{\Phi} \bigg[ \prod_{\substack{x_{i} \in \Phi \\ x_{i} > M}} \frac{1}{1 + s \widehat{\rho} r_{x_{i}}^{-\alpha}} \prod_{x_{i} \in \Phi} \frac{1}{1 + s \widetilde{\rho} r_{\widetilde{m}_{x_{i}}}^{-\alpha}} \bigg] = \Exp_{\Phi} \bigg[ \prod_{\substack{x_{i} \in \Phi \\ x_{i} \leq M}} \frac{1}{1 + s \widehat{\rho} r_{\widetilde{m}_{x_{i}}}^{-\alpha}} \prod_{\substack{x_{i} \in \Phi \\ x_{i} > M}} \frac{1}{1 + s \widehat{\rho} r_{x_{i}}^{-\alpha}} \frac{1}{1 + s \widetilde{\rho} r_{\widetilde{m}_{x_{i}}}^{-\alpha}} \bigg].
\end{align}
Finally, the two products in the expectation on the right-hand side of \eqref{eq:L_I1_M1} are independent and can be thus separated. \hfill \IEEEQED

\subsection{Proof of Corollary~\ref{cor:canc_UL}} \label{sec:A_P_suc1_canc1}

The average \red{SI} power is larger than the average interference power from the nearest FD BS~if
\begin{align} \label{eq:canc_UL1}
\frac{1}{(1 + s b \widehat{\rho})^{a}} \leq \Exp \bigg[ \frac{1}{1 + s \widehat{\rho} r_{x_{1}}^{-\alpha}} \bigg]
\end{align}
where
\begin{align}
\nonumber \Exp \bigg[ \frac{1}{1 + s \widehat{\rho} r_{x_{1}}^{-\alpha}} \bigg] & = 2 \pi \lambda \int_{0}^{\infty} \frac{1}{1 + s \widehat{\rho} r^{-\alpha}} e^{- \pi \lambda r^{2}} r \diff r \\
\label{eq:canc_UL2} & \simeq \exp \bigg( - 2 \pi \lambda \int_{0}^{d_{1}} \bigg( 1 - \frac{1}{1 + s \widehat{\rho} r^{-\alpha}} \bigg) r \diff r \bigg).
\end{align}
\red{Note that, unfortunately, a closed-form solution for the first line in \eqref{eq:canc_UL2} is not available and thus we resort to \cite{Atz16b} to obtain the approximation in the second line in \eqref{eq:canc_UL2}, with $d_{1}$ representing the average distance to the nearest FD BS (cf. \eqref{eq:d_M}). Since $d_{1} = \frac{1}{2 \sqrt{\lambda}}$ \cite[Ch.~2.9.1]{Hae12}, the condition in \eqref{eq:canc_UL} follows from solving the integral in the second line in \eqref{eq:canc_UL2} and plugging its solution into \eqref{eq:canc_UL1} with $s = \theta \widetilde{\rho}^{-1} \widetilde{R}^{\alpha}$.} \hfill \IEEEQED

\subsection{Proof of Corollary~\ref{cor:canc_imp_CSI}} \label{sec:A_P_suc1_canc2}

\red{Assume PZF is adopted to cancel the SI with imperfect SI channel estimation $\hat{\H}_{x x}$. The receive combining vector reads as}
\begin{align}
\red{\v_{x} = \frac{ \big( \I_{N_{\rmR}} - \hat{\H}_{x x} \w_{x} (\hat{\H}_{x x} \w_{x})^{\sharp} \big) \h_{\widetilde{m}_{x} x}}{\| (\I_{N_{\rmR}} - \hat{\H}_{x x} \w_{x} (\hat{\H}_{x x} \w_{x})^{\sharp}) \h_{\widetilde{m}_{x} x} \|}}
\end{align}
\red{and since $\v_{x}^{\herm} \hat{\H}_{x x} \w_{x} = 0$, the SI power is given by $|\v_{x}^{\herm} \H_{x x} \w_{x}|^{2} = |\v_{x}^{\herm} \E \w_{x}|^{2}$. Let $\E \triangleq \big( (e_{i k})_{i=1}^{N_{\rmR}} \big)_{j=1}^{N_{\rmT}}$; building on Lemma~\ref{lem:SI}, we can write (cf. \eqref{eq:vHw1})}
\begin{align}
\red{\Exp \big[ |\v^{\herm} \E \w|^{2} \big] \leq \Exp \bigg[ \sum_{i=1}^{N_{\rmR}} \sum_{j=1}^{N_{\rmT}} |v_{i}|^{2} |h_{i j}|^{2} |w_{j}|^{2} \bigg] = \frac{1}{N_{\rmR} N_{\rmT}} \sum_{i=1}^{N_{\rmR}} \sum_{j=1}^{N_{\rmT}} \Exp \big[ |e_{i j}|^{2} \big] = \epsilon^{2}}
\end{align}
\red{where the upper bound comes from applying the Cauchy-Schwartz inequality.} \hfill \IEEEQED


\subsection{Proof of Corollary~\ref{cor:canc_DL1}} \label{sec:A_P_suc2_canc1}

The average inter-node interference power is larger than the average interference power from the nearest FD BS if
\begin{align}
\Psi(s,\widehat{R}) \leq \Exp \bigg[ \frac{1}{1 + s \widehat{\rho} r_{x_{1}}^{-\alpha}} \bigg].
\end{align}
Then, we use the upper bound of $\Psi(s,r)$ in \eqref{eq:Psi_bounds} and $s = \theta \widehat{\rho}^{-1} \widehat{R}^{\alpha}$, and the rest of the proof follows similar steps as in Appendix~\ref{sec:A_P_suc1_canc1}. \hfill \IEEEQED

\subsection{Proof of Corollary~\ref{cor:canc_DL2}} \label{sec:A_P_suc2_canc2}

Recall the definition of $\setL_{I_{\widehat{m}_{0}}} (s)$ in \eqref{eq:LI_2}. Considering the scenario where the FD BSs transmit with one antenna to HD DL nodes with two receive antennas, we can build on Theorem~\ref{th:P_suc2} to write the success probabilities of the second hop with and without inter-node interference~cancellation~as
\begin{align}
\mathrm{P}_{\mathrm{suc},1}^{(2)} (\theta) & \triangleq \exp \big( - \lambda \Upsilon(s) \big), \\
\mathrm{P}_{\mathrm{suc},2}^{(2)} (\theta) & \triangleq \bigg[ \setL_{I_{\widehat{m}_{0}}} (s) - s \frac{\diff}{\diff s} \setL_{I_{\widehat{m}_{0}}} (s) \bigg]_{s = \theta \widehat{\rho}^{-1} \widehat{R}^{\alpha}}
\end{align}
respectively. Now, let us fix $\widetilde{R} = \widehat{R} = R$ and let us consider the upper bound on $\Psi(s,r)$ in \eqref{eq:Psi_bounds} and $\Upsilon^{(\min)} (s)$ in \eqref{eq:Upsilon_min}. After some algebraic manipulations, we have that $\mathrm{P}_{\mathrm{suc},1}^{(2)} (\theta) \geq \mathrm{P}_{\mathrm{suc},2}^{(2)} (\theta)$ when the condition in \eqref{eq:canc_DL2} is satisfied. \hfill \IEEEQED

\addcontentsline{toc}{chapter}{References}
\bibliographystyle{IEEEtran}
\bibliography{IEEEabrv,ref_Huawei}

\end{document}